\documentclass[useAMS,usenatbib,usegraphicx]{mn2e}
\usepackage{verbatim}
\usepackage{xspace}
\usepackage{array}
\usepackage{aas_macros}
\usepackage{subfig}
\usepackage{amsmath}
\usepackage{amssymb}

\usepackage[total={17.8cm,24.0cm},centering]{geometry}
\usepackage{times,deluxetable}

\def\kms{km~s$^{-1}$\xspace}

\newcommand{\twc}{$^{12}$CO\xspace}
\newcommand{\thc}{$^{13}$CO\xspace}
\newcommand{\two}{$^{12}$CO(1-0)\xspace}
\newcommand{\twt}{$^{12}$CO(2-1)\xspace}
\newcommand{\tho}{$^{13}$CO(1-0)\xspace}
\newcommand{\tht}{$^{13}$CO(2-1)\xspace}
\newcommand{\hcopo}{HCO$^{+}$(1-0)\xspace}
\newcommand{\hcop}{HCO$^{+}$\xspace}
\newcommand{\hcno}{HCN(1-0)\xspace}
\newcommand{\Rone}{$\mathfrak{R}_{10}$\xspace}
\newcommand{\Rtwo}{$\mathfrak{R}_{21}$\xspace}
\newcommand{\atlas}{ATLAS$^{\rm 3D}$\xspace}
\newcommand{\hi}{H$\:${\small I}\xspace}
\newcommand{\oiii}{[O$\:${\small III}]\xspace}
\newcommand{\nii}{[N$\:${\small II}]\xspace}
\newcommand{\sii}{[S$\:${\small II}]\xspace}
\newcommand{\hbeta}{H$\beta$\xspace}
\newcommand{\halpha}{H$\alpha$\xspace}

\title[The ATLAS$^{\;\rm 3D}$\xspace Project -- XI. Dense molecular gas]{The \atlas Project -- XI. Dense molecular gas properties of CO-luminous early-type galaxies\thanks{Based on observations carried out with the IRAM 30-m telescope. IRAM is supported by INSU/CNRS (France), MPG (Germany) and IGN (Spain).}}

\author
[Alison Crocker et al.]{\parbox{\textwidth}{Alison Crocker,$^{1}$\thanks{E-mail: \texttt{crocker@astro.umass.edu}}
Melanie Krips,$^{2}$
Martin Bureau,$^{3}$
Lisa M. Young,$^{4}$
Timothy A. Davis,$^{3}$
Estelle Bayet,$^{3}$
Katherine Alatalo,$^{5}$
Leo Blitz,$^{5}$
Maxime Bois,$^{6,7}$
Fr\'ed\'eric Bournaud,$^{8}$
Michele Cappellari,$^{3}$
Roger L. Davies,$^{3}$
P. T. de Zeeuw,$^{6,9}$
Pierre-Alain Duc,$^{8}$
Eric Emsellem,$^{6,7}$
Sadegh Khochfar,$^{10}$
Davor Krajnovi\'c,$^{6}$
Harald Kuntschner,$^{11}$
Pierre-Yves Lablanche,$^{6,7}$
Richard M. McDermid,$^{12}$
Raffaella Morganti,$^{13,14}$
Thorsten Naab,$^{15}$
Tom Oosterloo,$^{13,14}$
Marc Sarzi,$^{16}$
Nicholas Scott,$^{17}$
Paolo Serra,$^{13}$
and Anne-Marie Weijmans,$^{18}$\thanks{Dunlap Fellow}}\vspace{0.4cm}\\ 
\parbox{\textwidth}{$^{1}$Department of Astronomy, University of Massachusetts, Amherst, MA, 01003, USA\\
$^{2}$Institut de Radio Astronomie Millim\'etrique (IRAM), Domaine Universitaire, 300 rue de la Piscine, 38406 Saint Martin d'H\`eres, France\\
$^{3}$Sub-department of Astrophysics, Department of Physics, University of Oxford, Denys Wilkinson Building, Keble Road, Oxford, OX1 3RH, UK\\
$^{4}$Physics Department, New Mexico Institute of Mining and Technology, Socorro, NM 87801, USA\\
$^{5}$Department of Astronomy, Campbell Hall, University of California, Berkeley, CA 94720, USA\\
$^{6}$European Southern Observatory, Karl-Schwarzschild-Str. 2, 85748 Garching, Germany\\
$^{7}$Universit\'e Lyon 1, Observatoire de Lyon, Centre de Recherche Astrophysique de Lyon and Ecole Normale Sup\'erieure de Lyon, 9 avenue Charles Andr\'e, F-69230 Saint-Genis Laval, France\\
$^{8}$Laboratoire AIM Paris-Saclay, CEA/IRFU/SAp -- CNRS -- Universit\'e Paris Diderot, 91191 Gif-sur-Yvette Cedex, France\\
$^{9}$Sterrewacht Leiden, Leiden University, Postbus 9513, 2300 RA Leiden, the Netherlands\\
$^{10}$Max-Planck Institut f\"ur Extraterrestrische Physik, PO Box 1312, D-85478 Garching, Germany\\
$^{11}$Space Telescope European Coordinating Facility, European Southern Observatory, Karl-Schwarzschild-Str. 2, 85748 Garching, Germany\\
$^{12}$Gemini Observatory, Northern Operations Centre, 670 N. A`ohoku Place, Hilo, HI 96720, USA\\
$^{13}$Netherlands Institute for Radio Astronomy (ASTRON), Postbus 2, 7990 AA Dwingeloo, The Netherlands\\
$^{14}$Kapteyn Astronomical Institute, University of Groningen, Postbus 800, 9700 AV Groningen, The Netherlands\\
$^{15}$Max-Planck-Institut f\"ur Astrophysik, Karl-Schwarzschild-Str. 1, 85741 Garching, Germany\\
$^{16}$Centre for Astrophysics Research, University of Hertfordshire, Hatfield, Herts AL1 9AB, UK\\
$^{17}$Centre for Astrophysics \& Supercomputing, Swinburne University of Technology, PO BOX 218, Hawthorn, VIC 3122, Australia\\
$^{18}$Dunlap Institute for Astronomy \& Astrophysics, University of Toronto, 50 St. George Street, Toronto, ON M5S 3H4, Canada\\
}}

\date{}

\pagerange{\pageref{firstpage}--\pageref{lastpage}} \pubyear{2011}

\begin{document}
\label{firstpage}
\maketitle
\clearpage

\begin{abstract}   
Surveying eighteen \twc-bright galaxies from the \atlas early-type galaxy sample with the Institut de Radio Astronomie Millim\'etrique (IRAM) 30m telescope, we detect \tho and \tht in all eighteen galaxies, \hcno in 12/18 and \hcopo in 10/18. We find that the line ratios \two/\tho and \two/\hcno are clearly correlated with several galaxy properties: total stellar mass, luminosity-weighted mean stellar age, molecular to atomic gas ratio, dust temperature and dust morphology. We suggest that these correlations are primarily governed by the optical depth in the \twc lines; interacting, accreting and/or starbursting early-type galaxies have more optically thin molecular gas while those with settled dust and gas discs host optically thick molecular gas. The ranges of the integrated line intensity ratios generally overlap with those of spirals, although we note some outliers in the \two/\tho, \twt/\tht and HCN/\hcopo ratios.  In particular, three galaxies are found to have very low \two/\tho and \twt/\tht ratios. Such low ratios may signal particularly stable molecular gas which creates stars less efficiently than `normal' (i.e. below Schmidt-Kennicutt prediction), consistent with the low dust temperatures seen in these galaxies. 

\end{abstract}

\begin{keywords}
galaxies: elliptical and lenticular,
cD -- galaxies: ISM -- galaxies: stellar content -- galaxies:
evolution -- galaxies: kinematics and dynamics  
\end{keywords}

\section{Introduction}
Molecular gas is an essential ingredient for star formation found in many, but not all, galaxies. Early-type galaxies (ellipticals and lenticulars) were classically thought to completely lack molecular gas and to be passively-evolving `red and dead' galaxies. However, we have long known that not all early-types are empty of cold gas; molecular gas was first detected in early-type galaxies by \citet{wiklind86} and \citet{phillips87}. 

Shortly after these first detections, surveys detected 10-20 galaxies, but were biased towards early-types with particular properties, such as those bright in the far infra-red \citep{wiklind89,sage89} or with optically obscuring dust \citep{wang92}. These selection effects made it easy to dismiss any early-type galaxies with molecular gas as peculiar systems. The next generation of surveys increased numbers (30-50 galaxies) and were less biased, if still not complete \citep{welch03,sage07,combes07}. In \citet[][hereafter Paper IV]{young11}, we have recently completed an extensive (259 galaxies) molecular gas detection campaign for the volume-limited \atlas sample of early-type galaxies \citep[][hereafter Paper I]{cappellari11a}. We find a 22\% detection rate, down to a typical detection threshold of $6 \times 10^7$ M$_{\sun}$ of H$_{2}$, robustly establishing that many early-type galaxies host a substantial amount of molecular gas.

The majority of early-type galaxies with molecular gas are obviously star-forming, based on ultraviolet (UV), optical or infrared data. Indeed, the detection rate for star formation seen in UV nearly matches the molecular detection rate ($\approx30\%$, \citealp{kaviraj07}). According to the current observations, the star formation efficiencies of early-type galaxies broadly follow the Kennicutt-Schmidt law \citep{shapiro10, crocker11}. A subset shows no obvious sign of ongoing star formation \citep{crocker08,crocker11}, although these determinations are difficult due to their very low specific star formation rates. Recently, \citet{saintonge11b} have determined that star formation efficiencies are reduced for more massive, more concentrated and higher stellar surface density galaxies, all properties that positively correlate with early-type morphology.  

In a spiral galaxy, the presence of a stellar disc renders the gas disc more locally unstable to axisymmetric perturbations, likely boosting its star formation efficiency \citep[e.g.][]{jog84}. In \citet[][hereafter Paper II]{krajnovic11} and \citet[][hereafter Paper III]{emsellem11}, we found that the vast majority of early-type galaxies (the fast rotators) are consistent with being a family of disc-like systems resembling spiral galaxies with the gas and dust removed \citep[][hereafter Paper VII]{cappellari11b}. However, the fast rotators are generally characterized by larger spheroids than spiral galaxies. This increase in the depth of the potential well is expected to make their gas discs more stable against fragmentation \citep{kawata07}. 

Indeed, simulations with a centrally-concentrated stellar mass distribution (as found in spheroids) and no stellar disc show that the cool gas is more stable than in spiral galaxies \citep{martig09}. This stability (termed `morphological quenching')  should lower the efficiency of star formation and produce a cool interstellar medium (ISM) with properties (velocity dispersion, density distribution, etc.) different from those of galaxies with stellar discs. In this paper, we present the first major attempt at constraining the empirical properties of the molecular gas in early-type galaxies, especially looking for any divergence from the properties found for spiral galaxies. 

The surveys mentioned above have used the bright \two emission line to measure the total molecular content of early-type galaxies, but little work has been done to constrain the molecular gas properties using other species and transitions.  Several other molecular species are bright enough to measure, including \thc, HCN and HCO$^{+}$. These species have been widely observed in starburst and Seyfert galaxies and also in some local spiral galaxies. 

The observed \twc/\thc ratio is widely used to indicate the average optical depth of the molecular gas\footnote{Described by: $(\frac{^{12}\mathrm{CO}}{^{13}\mathrm{CO}})_{\mathrm {obs}}=\frac{e^{-\tau(12)}}{e^{-\tau(13)}}\frac{^{12}\mathrm{CO}}{^{13}\mathrm{CO}}$}, although it may also be influenced by chemical processes. The \twc isotope is far more abundant and becomes optically thick at lower H$_{2}$ column densities. Its use as a measure of the total molecular hydrogen content (via one of the \twc-to-H$_{2}$ conversion factors X$_{\mathrm {CO}}$ or $\alpha_{\mathrm {CO}}$) relies on the assumption that it is optically thick, and essentially counts the number of virialized molecular clouds of a similar temperature and density \citep[e.g.][]{young91}. The less abundant \thc isotope is optically thin until higher column densities and thus the \twc/\thc ratio (often denoted $\mathfrak{R}$) should reflect differences in average optical depth in \twc, either within the molecular clouds themselves or because of the additional contribution of a diffuse molecular component. Variations in the $^{12}$C to $^{13}$C abundance ratio and \twc$\leftrightarrow$\thc fractionation (due to charge-ion reactions or selective photodissociation) must also be considered, although they do not seem to drive this ratio in spiral galaxies \citep{paglione01}. 

The \hcno and \hcopo lines have higher critical densities (approximately $10^{6}$ and $10^{5}$ cm$^{-3}$, respectively) than CO(1-0) ($n_{\mathrm crit} \approx 10^{3}$)\footnote{Note that effective critical densities, which take into account radiative trapping, are about an order of magnitude lower \citep[e.g.][]{scoville74}}. The critical density is simply the density at which collisions are more frequent than radiative decays, not a strict limit with no emission at lower densities. Thus depending on the density distribution of the molecular gas, the typical gas density probed by the \hcno and \hcopo transitions will vary from system to system, although in all cases it will be higher than the densities probed by CO(1-0). 

Higher HCN/\twc and \hcop/\twc ratios in luminous infrared galaxies (LIRGs) and ultra-luminous infrared galaxies (ULIRGs) are thus taken to indicate higher dense gas fractions in these galaxies \citep{gao04a, gracia-carpio06}. The more moderate increase of \hcop/\twc compared to HCN/\twc in these systems may be explained by the order of magnitude difference in the critical densities of \hcno and \hcopo \citep{juneau09}. In both the Milky Way and M31, the HCN/\twc ratio declines with radius, signaling the decline of the dense gas content \citep{helfer97, brouillet05}. Indeed, \citet{helfer97} find that the \hcno/\two ratio correlates with the hydrostatic pressure in the disc: \hcno/\two $\propto P^{\,0.19\pm0.04}$.

Molecular chemistry can also influence the HCN and \hcop emission seen in galaxies. X-ray dominated regions (XDRs) around active galactic nuclei (AGN) may enhance the HCN abundance relative to CO \citep[e.g.][]{lepp96}. This effect has been observed in Seyfert galaxies and led to the recommendation to prefer \hcop as a dense gas tracer \citep{gracia-carpio06, krips07}. However, \citet{papadopoulos07} warn about possible effects of free electrons in cosmic-ray dominated regions (CRDRs) and XDRs, or highly turbulent molecular clouds destroying \hcop. Nevertheless, most of these chemical effects rely on the conditions found in starbursts or AGN, and hence HCN and \hcop are likely to remain good tracers of dense gas in more quiescent regimes. 

Until recently, the only early-type galaxy to be studied in molecular species other than \twc was Centaurus A (Cen A), with its striking lane of dust and gas. The $\approx5\times10^{8}$ M$_{\sun}$ of molecular gas in its disc extends to about 2.6~kpc \citep{phillips87,eckart90a}, while 4$\times10^{8}$ M$_{\sun}$ of atomic gas is seen in a warped disc out to about 6~kpc \citep{vangorkom90,struve10}. While often considered a galaxy completely in its own class, its molecular and atomic gas content is not very different to that of some of our sample galaxies. In terms of molecular emission line studies, \citet{wild97} found a \two/\tho ratio of 14 towards the center of Cen A and 11 towards two positions in its disc, similar values as found in spirals. As for HCN, \citet{wild00} find a higher ratio of \hcno/\two in the center (0.067) than in offset positions along the disc (0.02-0.04).

In \citet{krips10}, the \tho,\tht, \hcno and \hcopo emission lines of four early-type galaxies were measured as a pilot project for this work. Two of these galaxies have particularly low \two/\tho ratios of around 3 and two have more typical ratios of around 10. The \hcno/\tho and \hcno/\two ratios are in the range observed for spiral and Seyfert galaxies, but lower than for starbursts. Most curiously, none were detected in \hcopo, despite 3/4 being detected in \hcno, which usually has a similar integrated intensity. In one case, the \hcno/\hcopo ratio is constrained to be larger than 2, indicating some significant difference in the chemistry or physical properties of this galaxy's molecular gas compared to those of spiral, Seyfert and starburst galaxies. 

In this paper, we extend the original sample of 4 early-type galaxies from \citet{krips10} to a sample of 18 galaxies from the \atlas sample. Section~2 describes the sample selection and Section~3 describes our observations and data reduction. We compare the profiles of the different molecular lines within each galaxy and discuss the derived line ratios in Section~4.  In Section~4, we also investigate the variations of the molecular line ratios with other galaxy properties. Our conclusions are presented in Section~5.



\begin{figure}
\begin{center}
\includegraphics[height=4.5cm]{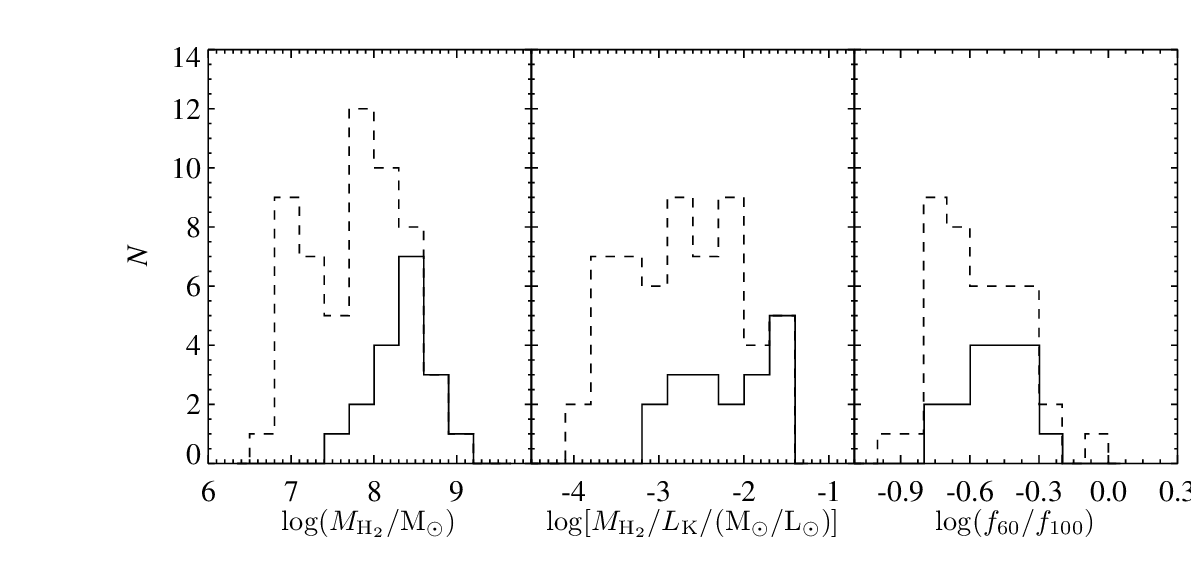}

\includegraphics[height=4.5cm]{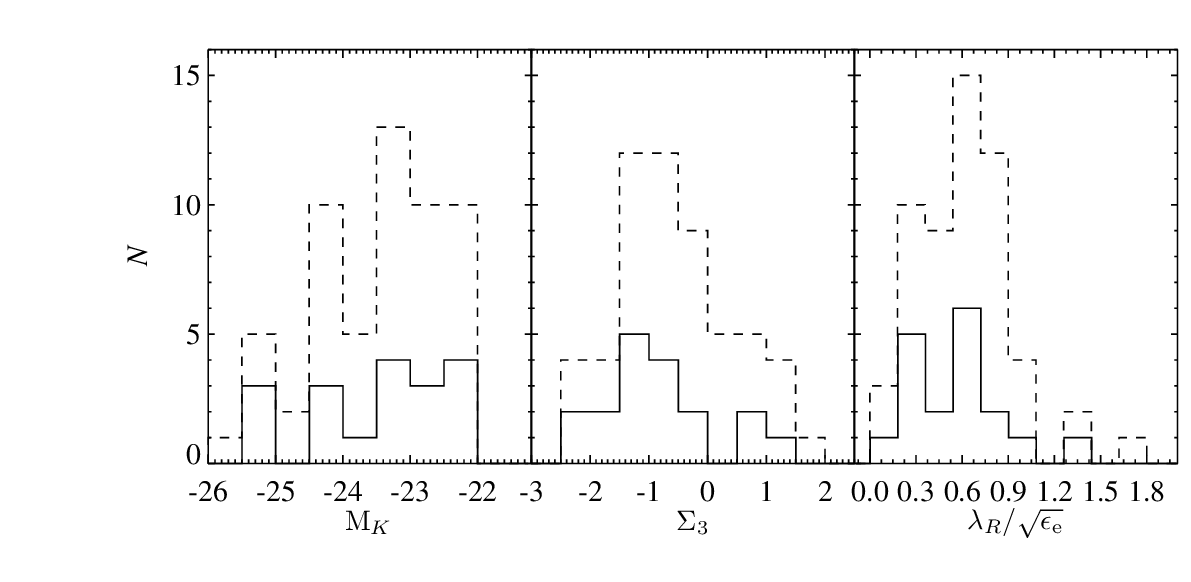}
\caption{Distributions of the 56 galaxies selected for follow-up (solid lines) out of the 59 \two detections from \atlas (dashed lines). Top panel: our subsample is clearly biased towards the highest molecular gas masses and mass fractions and slightly biased towards warmer dust temperatures, as measured by the Infrared Astronomical Satellite (IRAS) 60 to 100 \micron~flux ratio ($f_{60}/f_{100}$). The bottom panel shows no bias in absolute $K$-band magnitude, local environment ($\Sigma_{3}$), or specific stellar angular momentum ($\lambda_{\mathrm{R}}/\sqrt{\epsilon_{\mathrm e}}$).}
\label{fig:sample}
\end{center}
\end{figure}

\section{Sample}
We draw our sample from the \two detections of Paper IV which presents \two data on 259 of the 260 early-type galaxies in the \atlas
sample. The \atlas 
sample is volume-limited and complete to a distance $D<42$~Mpc, with a cut-off absolute magnitude of
$M_{K} = -21.5$. It includes only early-type galaxies, defined according to classic galaxy morphology \citep{hubble36,sandage61} by the absence of spiral arms (for non-edge on galaxies) or by the lack of galaxy-scale dust lanes (for edge on galaxies).
Further details on the sample can be found in
Paper I.  

Paper IV detects \twc in 56/259 of the \atlas sample
galaxies, based on centrally-pointed observations using the Institut de Radio Astronomie Millim\'etrique (IRAM) 30m telescope.  A rms noise of around 3.0 mK ($T_{\mathrm A}^{*}$) in \two was achieved for every galaxy. With a conversion factor X${_\mathrm{CO}=3.0 \times 10^{20}}$ cm$^{-2}$ (K km s$^{-1}$)$^{-1}$, this results in a molecular gas mass detection limit of  $1 \times 10^{7}$ M$_{\sun}$ for the nearest sample galaxies (11~Mpc) and $1 \times 10^{8}$ M$_{\sun}$ for the furthest sample galaxies (40~Mpc).
Due to the expected lower intensity of the \tho, \tht, HCN(1-0) and
\hcopo lines, we selected the 18 strongest \two detections for follow-up at
these transitions. Both peak brightness temperature and line width
were used to determine the `strength' of the \two detections. Out of
these 18 galaxies, four were previously observed and reported in the
\citet{krips10} pilot study.
Details of the sample galaxies are provided in Table~\ref{tab:sample},
including the H$_{2}$ mass derived in Paper IV. 

The necessity of choosing the brightest \two detections biases our sample towards galaxies with large molecular gas masses and large molecular gas mass fractions, as can be seen in the top panel of Fig.~\ref{fig:sample}. However, using Kolmogorov-Smirnov tests, the 18 galaxies selected are consistent with being a randomly drawn sub-sample of both the remainder of the CO-detected galaxies and the non-CO detected \atlas galaxies, with respect to the distribution of their absolute $K$-band magnitude, $\Sigma_{3}$ (a measure of local environment density; see Paper~VII) and $\lambda_{R}$  (a measure of the stellar specific angular momentum used to classify galaxies as fast or slow rotators; see Paper III). The distributions of these three parameters are plotted in the bottom panel of Fig.~\ref{fig:sample}. The selected subsample is slightly biased with respect to dust temperature (indicated by the $f_{60}/f_{100}$ ratio) when compared to the other CO detections (top panel of Fig.~\ref{fig:sample}). The probability that the two distributions are drawn from the same parent population is only 1.1\%, the CO-rich subsample being biased toward higher dust temperatures, but also possibly being more narrowly spread than the other CO detections.

\begin{table}
\caption{Observed line frequencies and beam sizes.}
\begin{center}
\begin{tabular}{lrr}
\hline
Line & Rest freq. & Beam size \\
 & (GHz)\phantom{} & (\arcsec)\phantom{} \\
 \hline
 \hcno &  88.632 & 27.7\\
 \hcopo & 	89.189 & 27.7\\
 \tho & 110.201 & 22.3\\
 \tht & 220.399 & 11.2\\
 \hline
 \end{tabular}
 \label{tab:lines}
 \end{center}
 \end{table}

\begin{table*}
\caption{Sample properties.}
\begin{center}
\begin{tabular}{lrrrrrrrrrr}
\hline
Name & D\phantom{c)} & M$_{K}\phantom{)}$ & $\lambda_{R_{\mathrm e}}$ & $\epsilon_{\mathrm e}\phantom{0}$ & $\rho_{10}$\phantom{11} & log(M$_{\mathrm{HI}}$)  & log(M$_{\mathrm{H_{2}}}$) & $f_{60}/f_{100}$ & Dust Morph. & \oiii/\hbeta\\ 
 & (Mpc) & (mag) &  & & (Mpc$^{-3}$) & (M$_{\sun}$) & (M$_{\sun}$) & &\\
 \hline
IC0676      & 24.6 & -22.27 & 0.49 & 0.524 &  -1.41 & 8.27$^{1}$ &  8.62 & 0.62 & F & 0.32\\
IC1024      & 24.2 & -21.85 & 0.72 & 0.679 & -0.85 & 9.04$^{2}$ & 8.57 & 0.52 & F\ & 0.60\\
NGC1222 & 33.3 & -22.71 & 0.15 & 0.280 & -1.90 & 9.31$^{2}$ &9.08 & 0.85 & F & 1.81\\
NGC1266 & 29.9 & -22.93 & 0.64 & 0.193 & -2.11 & 6.98{$^{3}$} & 9.29 & 0.79 & F & 1.70\\
NGC2764 & 39.6 & -23.19 & 0.66 & 0.614 & -2.01 & 9.28\phantom{$^{2}$} & 9.19 & 0.51& F & 0.44\\
NGC3032 & 21.4 & -22.01 & 0.34 & 0.102 & -1.52 & 8.04\phantom{$^{2}$} & 8.41 & 0.41 & D & 0.31\\
NGC3607 & 22.2 & -24.74 & 0.21 & 0.185 & -0.92 &  $<6.92$\phantom{$^{2}$} & 8.42 & -- & D & 0.99\\
NGC3665 & 33.1 & -24.94 & 0.41 & 0.216 & -1.94 & $<7.43$\phantom{$^{2}$} & 8.91 & 0.25 & D & 0.61\\
NGC4150 & 13.4 & -21.65 & 0.51 & 0.328 & -1.18 & 6.26\phantom{$^{2}$} & 7.82 & 0.46 & N & 1.53\\
NGC4459 & 16.1 & -23.89 & 0.44 & 0.148 & 0.78 & $<6.91$\phantom{$^{2}$} & 8.24& 0.39 & D & 0.66\\
NGC4526 & 16.4 & -24.62 & 0.45 & 0.361 & 0.83 & $<7.68^{4}$ & 8.59 & 0.33 & D & 0.81\\
NGC4694 & 16.5 & -22.15 & 0.29 & 0.547 & 0.60 & 8.21\phantom{$^{2}$} & 7.99 & 0.41 & F & 0.68\\
NGC4710 & 16.5 & -23.53 & 0.65 & 0.699 & 0.09 & 6.84\phantom{$^{2}$} & 8.69 & 0.39 & D & 1.13\\
NGC5866 & 14.9 & -24.00 & 0.32 & 0.566 & -2.12 & 6.96\phantom{$^{2}$} & 8.47 & 0.27 & D & --\\
NGC6014 & 35.8 & -22.99 & 0.39 & 0.419 & -2.27 & $<8.28^{1}$ & 8.80 & 0.63 & D & 0.45\\
NGC7465 & 29.3 & -22.82 & 0.28 & 0.364 & -1.97 & 9.98\phantom{$^{2}$} & 8.80 & 0.47 & F & 0.87\\
PGC058114 & 23.8 & -21.57 & 0.18 & 0.185 & -2.31 & 8.23$^{5}$ & 8.61 & 0.78 & -- & 0.64\\
UGC09519 & 27.6 & -21.98 & 0.63 & 0.484 & -2.41 & 9.27\phantom{$^{2}$} & 8.80 & 0.40 & F & 1.14\\
\hline
\end{tabular}\\
\vspace{0.1cm}
Notes: Distances and M$_{K}$ from Paper I; $\lambda_{R_{\mathrm e}}$ and $\epsilon_{\mathrm e}$ from Paper III; $\rho_{10}$ from Paper VII; H$\,${\scriptsize I}\ content from \citet{serra11}; H$_2$ content from Paper IV; $f_{60}/f_{100}$ derived from IRAS measurements; dust morphology is D for disc, F for filamentary and N for none from Paper II. Exceptions: $^{1}$ \citet{grossi09}, $^{2}$ \citet{paturel03}, $^{3}$ \citet{alatalo11}, $^{4}$ \citet{knapp79}, $^{5}$ \citet{springob05}.
\end{center}
\label{tab:sample}
\end{table*}%

\section{IRAM 30m observations and data reduction}

We used the IRAM 30m telescope at Pico Veleta, Spain, to observe the
\tho, \tht, HCN(1-0) and \hcopo transitions in July 2009 and April
2010. The Eight Mixer Receiver (EMIR) was used, tuned to either the
\tho or HCN(1-0) 
redshifted frequency at 3mm and simultaneously to \tht at 1mm for both
3mm setups. The rest frequencies and beam sizes for each of the lines are specified in Table~\ref{tab:lines}. We observed in wobbler-switching mode with reference
position offsets of $\pm 120$\arcsec. The Wideband Line Multiple
Autocorrelator (WILMA) backend provided a bandwidth of 8~GHz at 2~MHz
channel resolution (2~MHz corresponds to 6.7, 5.5 and 2.7 \kms at
\tho, HCN(1-0) and \tht, respectively).  This bandwidth is sufficient to observe \hcopo and HCN(1-0) simultaneously. 

System temperatures ranged from 90 to 200 K at 3mm and 250 to 600 K at 1mm. The pointing was checked approximately every 2-3 hours or more frequently during periods of high wind. Focusing was performed at the beginning of the night, after sunrise or sunset, and/or after about 5-6 hours of observing. 

Observations for four sample galaxies (NGC~3032, NGC~4150, NGC~4459 and NGC~4526) were taken in August 2008 and published in \citet{krips10}. We note that these observations used the old SIS receiver and thus a different observing
setup than the other 14 galaxies. Observing details are described in
\citet{krips10}. 

Conversions from antenna temperature ($T_{\mathrm A}^{*}$) to main beam temperature were calculated by dividing by the ratio of the beam and forward efficiencies, $\eta \equiv B_{\mathrm {eff}}/F_{\mathrm {eff}}$. We used values appropriate for when the observations were taken. For the new observations, values tabulated in the EMIR Commissioning Report\footnote{http://www.iram.es/IRAMES/mainWiki/EmirforAstronomers} were used. To obtain $B_{\mathrm {eff}}$ and $F_{\mathrm {eff}}$ values at frequencies not specified in these tables, we linearly interpolated between the two nearest values. The conversion ratios used for the new observations are: $\eta$(HCN)=$\eta$(\hcop)=0.85, $\eta$(\tho)=0.83 and $\eta$(\tht)=0.66. For the older observations, we base our conversions on the measurements performed in August 2007 and June 2008. For the 1mm values, linear interpolation is again used. However, noting the discrepant values found for the C150 receiver, we follow the approach of Paper IV for the 3mm values. The 3mm  $B_{\mathrm {eff}}$ values are based on the aperture efficiency plot on p. 15 of the IRAM 2007 Annual Report\footnote{http://www.iram.fr/IRAMFR/ARN/AnnualReports/IRAM\_2007.pdf} ($B_{\mathrm {eff}}$=$1.21A_{\mathrm {eff}}$, where $A_{\mathrm {eff}}$ is the aperture efficiency). This results in: $\eta$(HCN)=$\eta$(\hcop)=0.79, $\eta$(\tho)=0.78, $\eta$(\two)=0.78, $\eta$(\tht)=0.65 and $\eta$(\twt)=0.63.


Linear baselines were fit to frequencies outside of the expected line region for each scan, then the scans were averaged together weighted by their system temperature. The spectra for four sample galaxies are shown in Fig.~\ref{fig:spectra} in the $T_{\mathrm{mb}}$ scale, with the rest in Appendix~1. The figures include \two and \twt spectra from Paper IV and \citet{welch03} for comparison.

As all of the galaxies have previous well measured molecular
linewidths from the \two data, we use this prior knowledge to help
ascertain detections and measure integrated intensities. 
In order for a galaxy to be detected, we require the integrated
intensity over its \two-detected velocity range to be three times greater than
the uncertainty expected over such a velocity width.  As in
\citet{sage07} and Paper IV, the
statistical uncertainty $\sigma_{\mathrm{I}}$ in a sum over
$N_{\mathrm l}$ channels of width $\delta v$ and rms noise level
$\sigma$ is $\sigma_{\mathrm{I}}^{2}=(\delta
v)^{2}\sigma^{2}N_{\mathrm l}(1+N_{\mathrm l}/N_{\mathrm b})$, where $N_{\mathrm b}$ is the number of baseline channels used (and thus the
$N_{\mathrm l}/N_{\mathrm  b}$ term contributes the uncertainty from
estimating the baseline level). The rms noise is measured outside of
the range where line emission is expected. If a galaxy is not
detected, we give three times the measured uncertainty as an upper
limit. Integrated intensities with errors are tabulated in Table~\ref{tab:ii}, along with the velocity range integrated over. We note that only the measurement errors are tabulated, systematic uncertainties are also important at around 10\%. 

\begin{table*}
\caption{\two, \twt, \tho, \tht, HCN(1-0) and \hcopo integrated intensities.}
\begin{center}
\begin{tabular}{lrrrrrrr}
\hline
Galaxy & Vel. range & \two & \twt & \tho & \tht & HCN(1-0) & \hcopo \\
 & (km s$^{-1})$ & (K km s$^{-1}$) & (K km s$^{-1}$) &(K km s$^{-1}$) &(K km s$^{-1}$) &(K km s$^{-1}$) &(K km s$^{-1}$) \\
\hline

   IC0676 & 1310--1516 & $ 11.46\pm  0.39$ & $ 16.78\pm  0.32$ & $  1.44\pm  0.11$ & $  1.97\pm  0.15$ & $  0.27\pm  0.06$ & $  0.27\pm  0.06$ \\
   IC1024 & 1359--1619 & $ 11.41\pm  0.38$ & $ 15.74\pm  0.36$ & $  0.72\pm  0.08$ & $  1.90\pm  0.12$ & $ <  0.27$ & $ <  0.28$ \\
  NGC1222 & 2253--2603 & $ 17.32\pm  0.45$ & $ 28.95\pm  0.49$ & $  0.78\pm  0.10$ & $  1.42\pm  0.20$ & $ <  0.36$ & $  0.37\pm  0.12$ \\
  NGC1266 & 1750--2500 & $ 34.76\pm  0.99$ & $105.22\pm  0.88$ & $  1.04\pm  0.09$ & $  4.04\pm  0.19$ & $  2.83\pm  0.15$ & $  1.97\pm  0.15$ \\
  NGC2764 & 2514--2944 & $ 16.17\pm  0.48$ & $ 24.54\pm  0.71$ & $  1.41\pm  0.10$ & $  2.01\pm  0.14$ & $  0.28\pm  0.08$ & $  0.47\pm  0.08$ \\
  NGC3032$^{1}$ & 1475--1630 & $  8.32\pm  0.23$ & $  6.61\pm  0.27$ & $  0.84\pm  0.07$ & $  1.23\pm  0.09$ & $  0.27\pm  0.04$ & $ <  0.33$ \\
   NGC3607 & 670--1227 & $ 10.44\pm  0.29$ & $ 19.54\pm  2.20$ & $  1.66\pm  0.17$ & $  2.96\pm  0.32$ & $  0.73\pm  0.10$ & $  0.51\pm  0.10$ \\
  NGC3665 & 1737--2432 & $ 11.97\pm  0.64$ & $ 14.35\pm  0.83$ & $  3.71\pm  0.19$ & $  5.64\pm  0.19$ & $  0.49\pm  0.08$ & $ <  0.23$ \\
     NGC4150$^{1}$ & 75--350 & $  6.04\pm  0.47$ & $ 10.99\pm  0.49$ & $  0.42\pm  0.07$ & $  0.87\pm  0.11$ & $ <  0.24$ & $ <  0.25$ \\
   NGC4459$^{1}$ & 980--1385 & $ 10.03\pm  0.56$ & $ 11.52\pm  0.51$ & $  2.99\pm  0.13$ & $  4.03\pm  0.21$ & $  0.59\pm  0.12$ & $ <  0.62$ \\
    NGC4526$^{1}$ & 280--980 & $ 21.63\pm  1.00$ & $ 32.26\pm  0.89$ & $  5.95\pm  0.28$ & $  6.76\pm  0.34$ & $  1.75\pm  0.12$ & $ <  0.63$ \\
  NGC4694 & 1083--1264 & $  6.14\pm  0.35$ & $  6.60\pm  0.28$ & $  0.35\pm  0.08$ & $  0.65\pm  0.14$ & $ <  0.20$ & $ <  0.20$ \\
   NGC4710 & 896--1368 & $ 31.65\pm  0.74$ & $ 40.42\pm  0.62$ & $  4.77\pm  0.15$ & $  6.13\pm  0.20$ & $  1.41\pm  0.10$ & $  0.89\pm  0.10$ \\
   NGC5866 & 432--1051 & $ 21.57\pm  0.38$ & $ 17.35\pm  0.68$ & $  3.27\pm  0.16$ & $  3.80\pm  0.18$ & $  0.84\pm  0.09$ & $  0.57\pm  0.09$ \\
  NGC6014 & 2266--2570 & $  7.48\pm  0.38$ & $ 10.81\pm  0.37$ & $  0.78\pm  0.13$ & $  1.54\pm  0.10$ & $  0.20\pm  0.05$ & $ <  0.14$ \\
  NGC7465 & 1827--2117 & $ 11.89\pm  0.39$ & $ 21.48\pm  0.43$ & $  0.63\pm  0.14$ & $  1.15\pm  0.06$ & $  0.13\pm  0.03$ & $  0.29\pm  0.03$ \\
PGC058114 & 1383--1768 & $ 11.37\pm  0.38$ & $ 21.65\pm  0.27$ & $  0.56\pm  0.09$ & $  1.45\pm  0.20$ & $ <  0.33$ & $  0.38\pm  0.11$ \\
 UGC09519 & 1501--1818 & $ 12.64\pm  0.38$ & $ 13.96\pm  0.29$ & $  0.45\pm  0.10$ & $  0.88\pm  0.18$ & $ <  0.21$ & $  0.29\pm  0.07$ \\
\hline
\end{tabular}\\
\vspace{0.1cm}

$^{1}$ Pilot sample galaxies from \citet{krips10}.
\end{center}
\label{tab:ii}
\end{table*}%

\begin{figure*}
\begin{center}
\includegraphics[height=13cm]{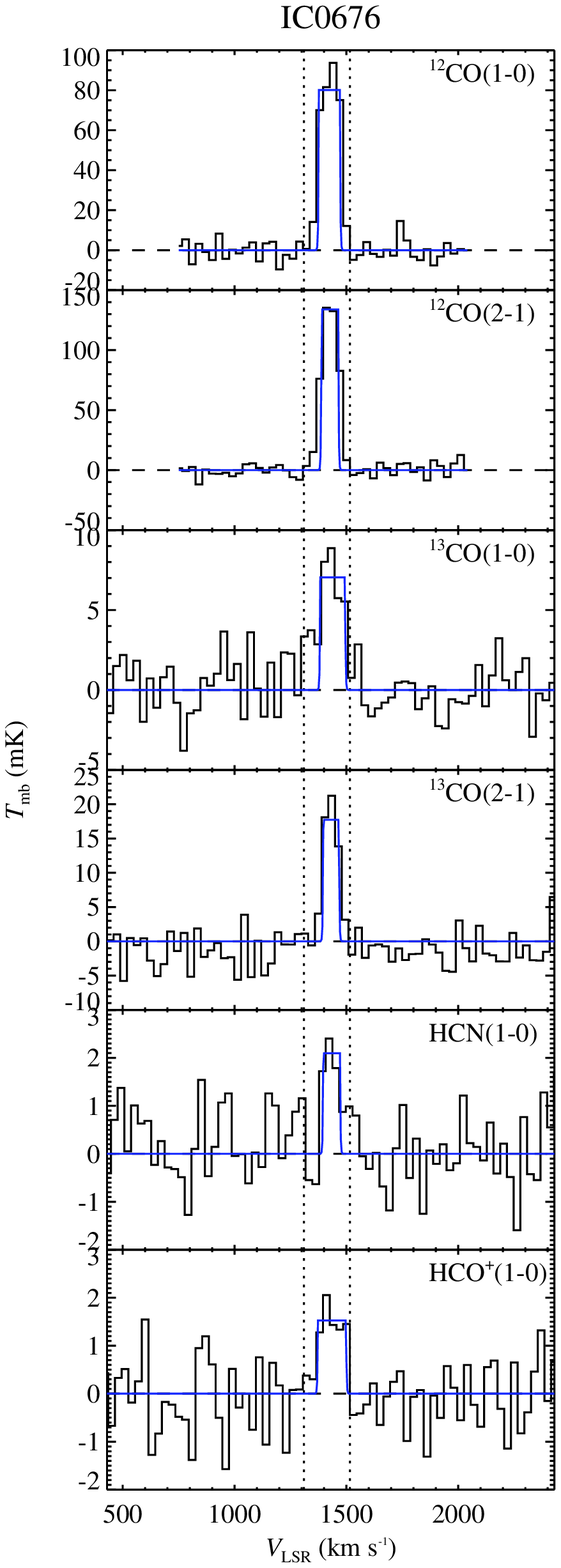}
\includegraphics[height=13cm]{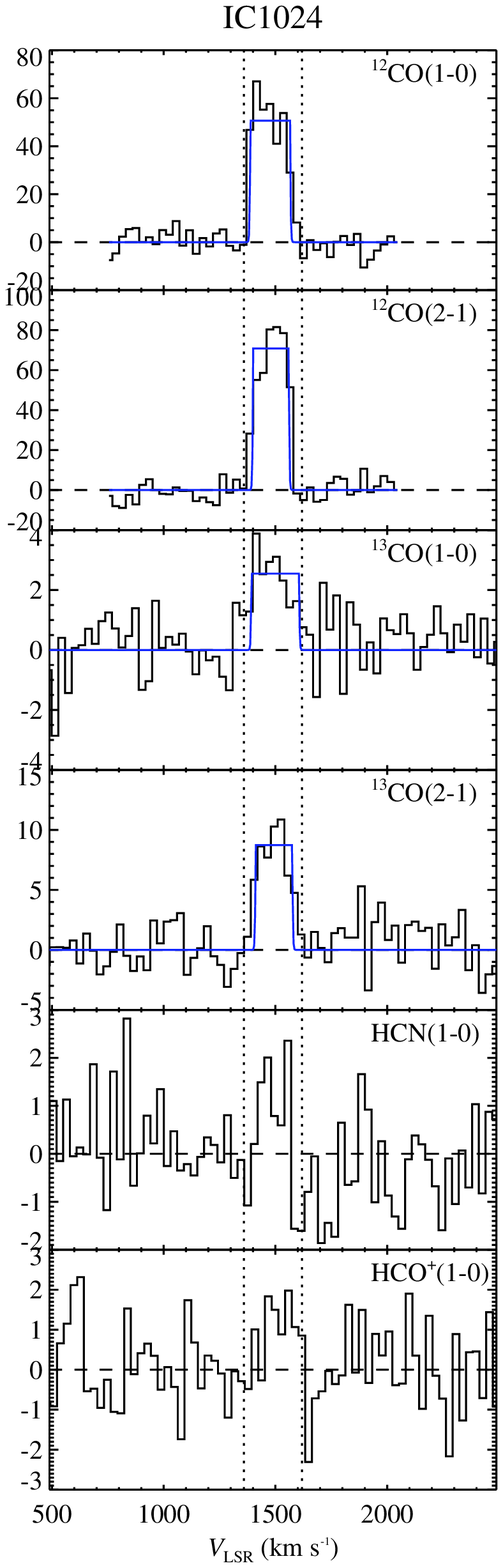}
\includegraphics[height=13cm]{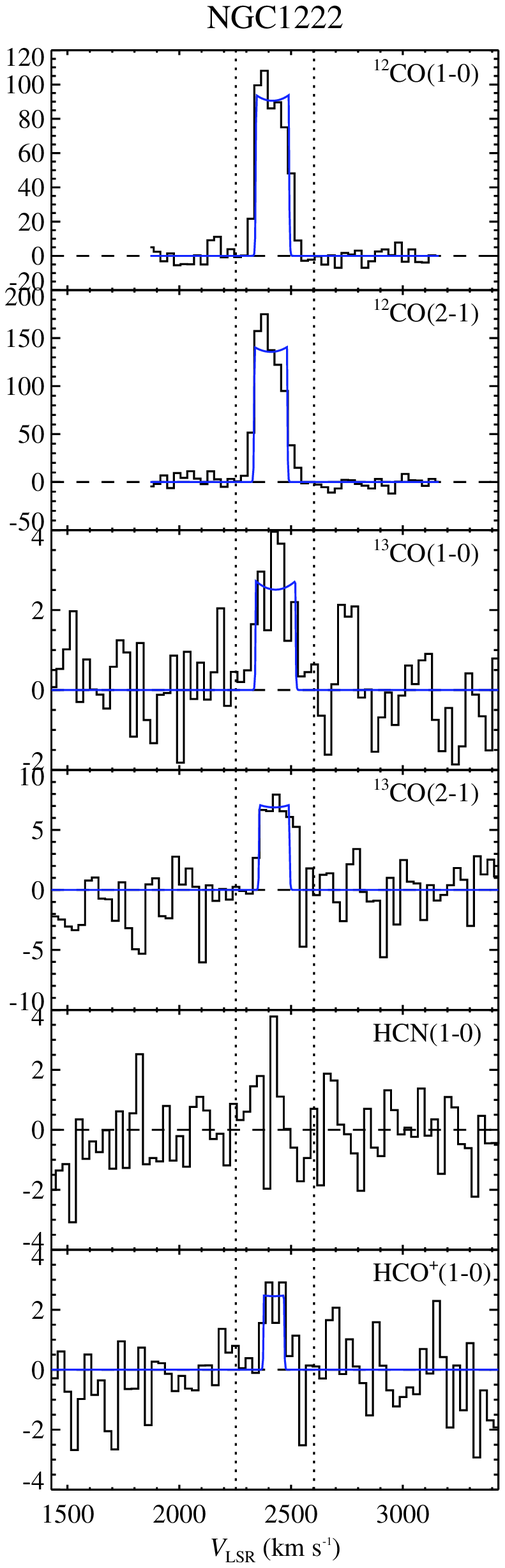}
\includegraphics[height=13cm]{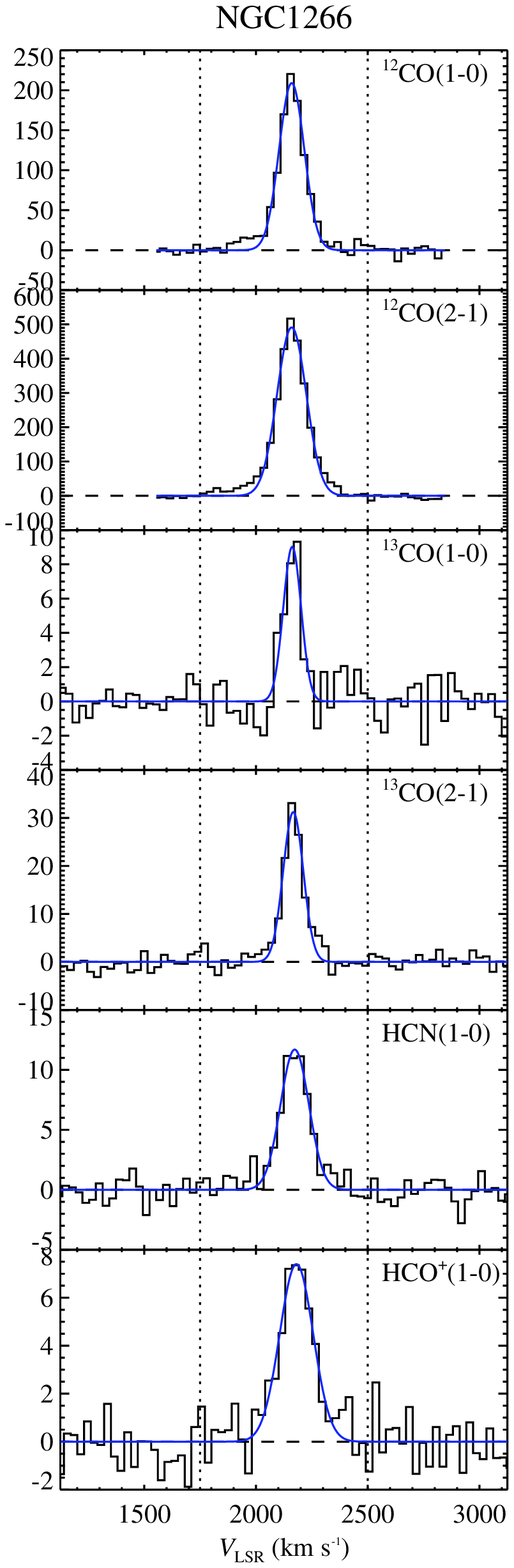}
\caption{Example molecular line spectra from the IRAM 30m telescope; spectra of the remaining 14 galaxies are shown in Appendix~A. The spectra have been binned to a channel width of 30 \kms
  and the scale is in main beam temperature (mK). The blue line shows the best-fit Gaussian or double-peak function (fits are only performed on detected lines). The vertical dashed lines indicate the velocity range integrated over to obtain integrated intensities. {\em Top to bottom:}
  \two, \twt, \tho, \tht, HCN(1-0) and \hcopo. The \two and \twt data are from
  Paper IV.}
\label{fig:spectra}
\end{center}
\end{figure*}

\begin{table*}
\caption{Integrated intensity line ratios.}
\begin{center}
\begin{tabular}{lrrrrrrr}
\hline
Galaxy  & $^{12}$CO/\thc & $^{12}$CO/\thc & HCN/\twc & HCN/\thc & HCN/\hcop & \thc/\hcop & \two/\\
	& (1-0)\phantom{000} & (2-1)\phantom{000} & (1-0)\phantom{000} & (1-0)\phantom{000} & (1-0)\phantom{000} & (1-0)\phantom{000} & \twt \phantom{}\\ 

\hline
\\[-0.25cm]

    IC0676 &   7.5 $^{+   1.3}_{-   1.2}$ &   8.3 $^{+   1.4}_{-   1.3}$ &  0.033 $^{+  0.008}_{-  0.008}$ &  0.25 $^{+  0.07}_{-  0.06}$ &   1.00 $^{+   0.35}_{-   0.26}$  &  4.03 $^{+   1.29}_{-   0.92  }$ & 1.34 $^{+   0.05}_{-   0.05}$\\ [0.1cm]
    IC1024 &  15.1 $^{+   3.0}_{-   2.6}$ &   8.1 $^{+   1.3}_{-   1.2}$ &  $<$  0.024 & $<$  0.38 &     ...  &$>$  2.61  & 1.17 $^{+   0.05}_{-   0.05}$\\ [0.1cm]
   NGC1222 &  21.1 $^{+   4.5}_{-   3.8}$ &  20.0 $^{+   4.5}_{-   3.7}$ &  $<$  0.021 & $<$  0.46 &  $<$   0.96  &  1.65 $^{+   0.86}_{-   0.49  }$ & 0.98 $^{+   0.03}_{-   0.03}$ \\ [0.1cm]
   NGC1266 &  30.9 $^{+   5.6}_{-   5.0}$ &  24.8 $^{+   3.8}_{-   3.6}$ &  0.128 $^{+  0.020}_{-  0.020}$ &  3.94 $^{+  0.73}_{-  0.66}$ &   1.43 $^{+   0.14}_{-   0.13}$  &  0.36 $^{+   0.07}_{-   0.07  }$ & 0.86 $^{+   0.03}_{-   0.03}$\\ [0.1cm]
   NGC2764 &  10.8 $^{+   1.8}_{-   1.7}$ &  11.6 $^{+   1.9}_{-   1.8}$ &  0.025 $^{+  0.008}_{-  0.008}$ &  0.27 $^{+  0.08}_{-  0.08}$ &   0.59 $^{+   0.21}_{-   0.17}$  &  2.22 $^{+   0.57}_{-   0.45  }$ & 1.50 $^{+   0.06}_{-   0.06}$ \\ [0.1cm]
   NGC3032 &   9.5 $^{+   1.6}_{-   1.5}$ &   5.3 $^{+   0.9}_{-   0.8}$ &  0.044 $^{+  0.009}_{-  0.009}$ &  0.41 $^{+  0.09}_{-  0.09}$ &  $>$   0.81  &$>$  2.52  & 2.04 $^{+   0.11}_{-   0.10}$ \\ [0.1cm]
   NGC3607 &   5.9 $^{+   1.1}_{-   1.0}$ &   6.4 $^{+   1.4}_{-   1.3}$ &  0.097 $^{+  0.019}_{-  0.019}$ &  0.58 $^{+  0.13}_{-  0.12}$ &   1.43 $^{+   0.40}_{-   0.29}$  &  2.47 $^{+   0.75}_{-   0.56  }$ & 1.03 $^{+   0.13}_{-   0.11}$ \\ [0.1cm]
   NGC3665 &   3.0 $^{+   0.5}_{-   0.5}$ &   2.4 $^{+   0.4}_{-   0.4}$ &  0.060 $^{+  0.013}_{-  0.013}$ &  0.18 $^{+  0.04}_{-  0.04}$ &  $>$   2.13  &$>$ 16.21 & 1.93 $^{+   0.16}_{-   0.15}$  \\ [0.1cm]
   NGC4150 &  13.7 $^{+   3.6}_{-   2.9}$ &  12.2 $^{+   2.7}_{-   2.2}$ &  $<$  0.040 & $<$  0.58 &     ...  &$>$  1.71  & 1.08 $^{+   0.10}_{-   0.09}$\\ [0.1cm]
   NGC4459 &   3.2 $^{+   0.5}_{-   0.5}$ &   2.8 $^{+   0.4}_{-   0.4}$ &  0.081 $^{+  0.020}_{-  0.020}$ &  0.26 $^{+  0.06}_{-  0.06}$ &  $>$   0.95  &$>$  4.80  & 1.39 $^{+   0.10}_{-   0.10}$ \\ [0.1cm]
   NGC4526 &   3.5 $^{+   0.5}_{-   0.5}$ &   4.6 $^{+   0.7}_{-   0.7}$ &  0.108 $^{+  0.018}_{-  0.017}$ &  0.38 $^{+  0.06}_{-  0.06}$ &  $>$   2.77  &$>$  9.41 & 1.15 $^{+   0.06}_{-   0.06}$ \\ [0.1cm]
   NGC4694 &  16.6 $^{+   5.7}_{-   3.9}$ &   9.7 $^{+   3.1}_{-   2.2}$ &  $<$  0.032 & $<$  0.56 &     ...  &$>$  1.73 & 1.87 $^{+   0.14}_{-   0.13}$ \\ [0.1cm]
   NGC4710 &   6.3 $^{+   0.9}_{-   0.9}$ &   6.5 $^{+   1.0}_{-   0.9}$ &  0.061 $^{+  0.010}_{-  0.010}$ &  0.39 $^{+  0.06}_{-  0.06}$ &   1.59 $^{+   0.24}_{-   0.20}$  &  4.10 $^{+   0.82}_{-   0.71  }$ & 1.32 $^{+   0.04}_{-   0.04}$ \\ [0.1cm]
   NGC5866 &   6.2 $^{+   0.9}_{-   0.9}$ &   4.3 $^{+   0.7}_{-   0.7}$ &  0.059 $^{+  0.010}_{-  0.010}$ &  0.36 $^{+  0.07}_{-  0.06}$ &   1.48 $^{+   0.32}_{-   0.25}$  &  4.10 $^{+   1.00}_{-   0.80  }$ & 3.07 $^{+   0.13}_{-   0.13}$\\ [0.1cm]
   NGC6014 &   9.0 $^{+   2.3}_{-   1.8}$ &   6.7 $^{+   1.1}_{-   1.1}$ &  0.040 $^{+  0.011}_{-  0.011}$ &  0.36 $^{+  0.13}_{-  0.11}$ &  $>$   1.46  &$>$  5.51 & 1.58 $^{+   0.10}_{-   0.09}$ \\ [0.1cm]
   NGC7465 &  17.9 $^{+   5.7}_{-   4.0}$ &  18.1 $^{+   2.8}_{-   2.7}$ &  0.015 $^{+  0.004}_{-  0.004}$ &  0.27 $^{+  0.11}_{-  0.09}$ &   0.44 $^{+   0.13}_{-   0.12}$  &  1.63 $^{+   0.48}_{-   0.44  }$ & 1.05 $^{+   0.04}_{-   0.04}$\\ [0.1cm]
  UGC09519 &  26.2 $^{+   8.9}_{-   6.0}$ &  15.1 $^{+   4.6}_{-   3.2}$ &  $<$  0.017 & $<$  0.47 &  $<$   0.72  &  1.08 $^{+   0.47}_{-   0.34  }$ & 1.18 $^{+   0.04}_{-   0.04}$ \\ [0.1cm]
 PGC058114 &  18.9 $^{+   4.6}_{-   3.6}$ &  14.4 $^{+   3.1}_{-   2.6}$ &  $<$  0.029 & $<$  0.58 &  $<$   0.85  &  1.05 $^{+   0.49}_{-   0.31  }$ & 2.30 $^{+   0.08}_{-   0.08}$\\ [0.1cm]

\hline

\end{tabular}
\end{center}
\label{tab:ratios}
\end{table*}%

%

\begin{figure}
\begin{center}
\includegraphics[width=7.5cm]{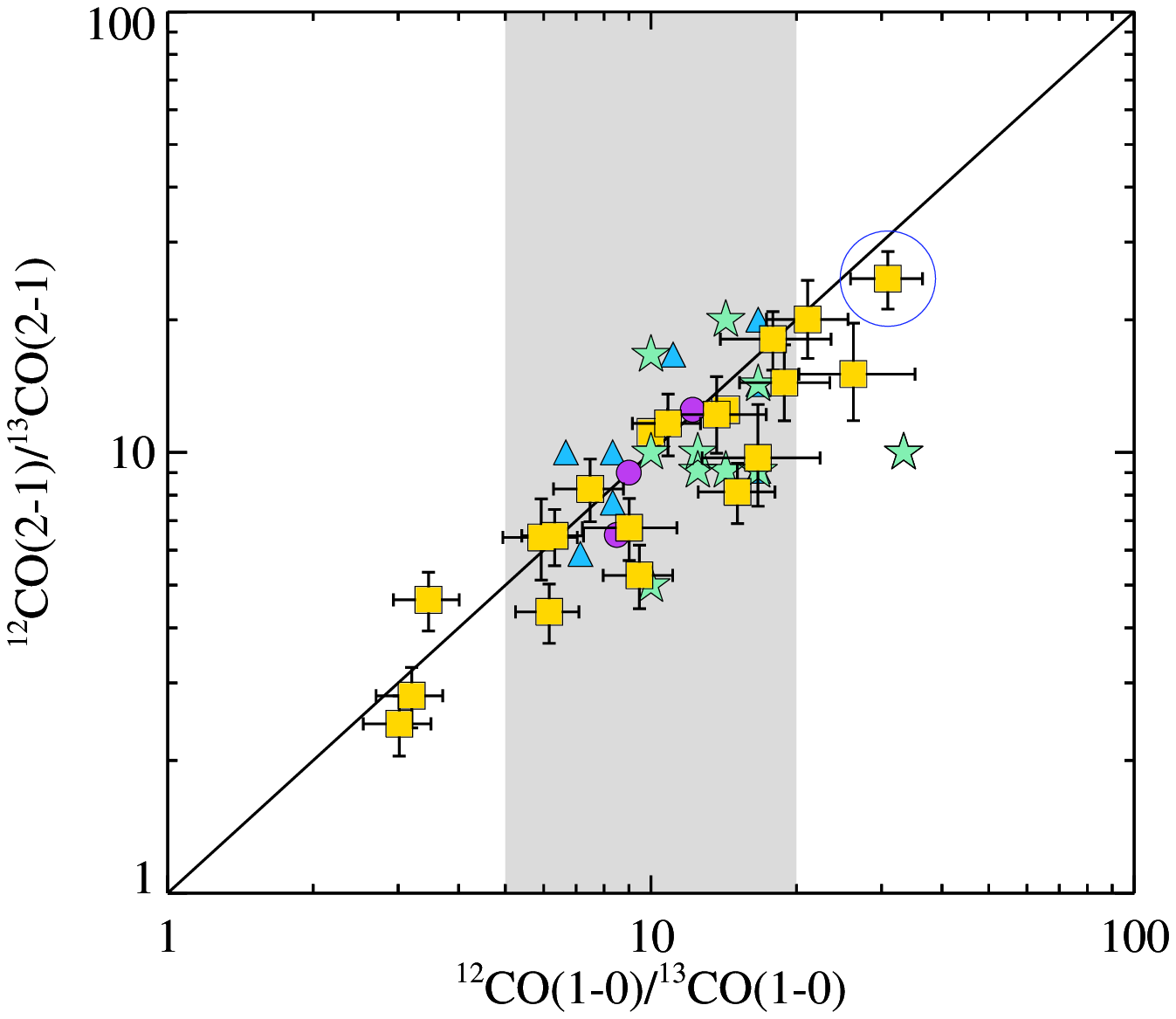}
\\
\includegraphics[width=7.5cm]{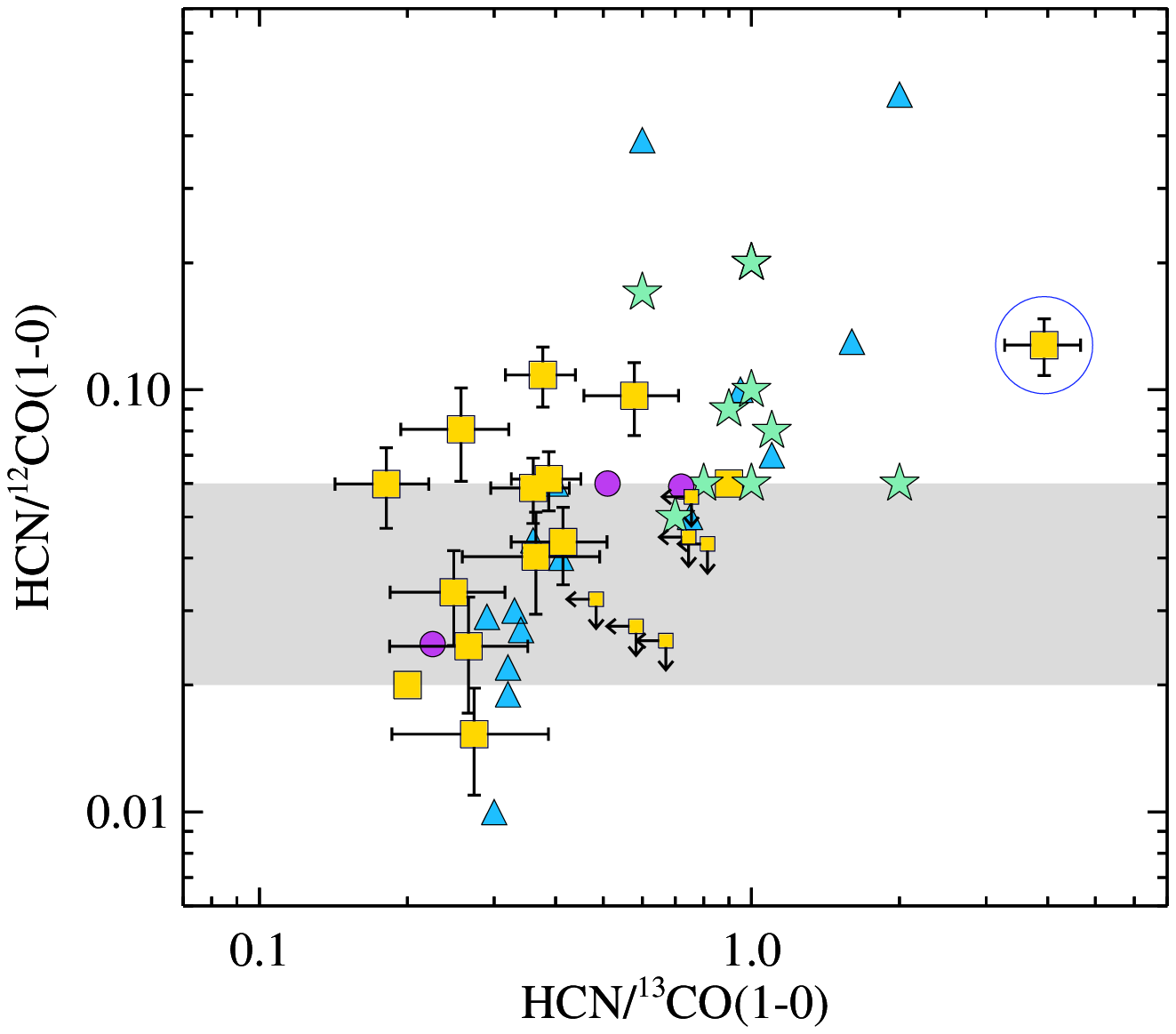}
\\
\includegraphics[width=7.5cm]{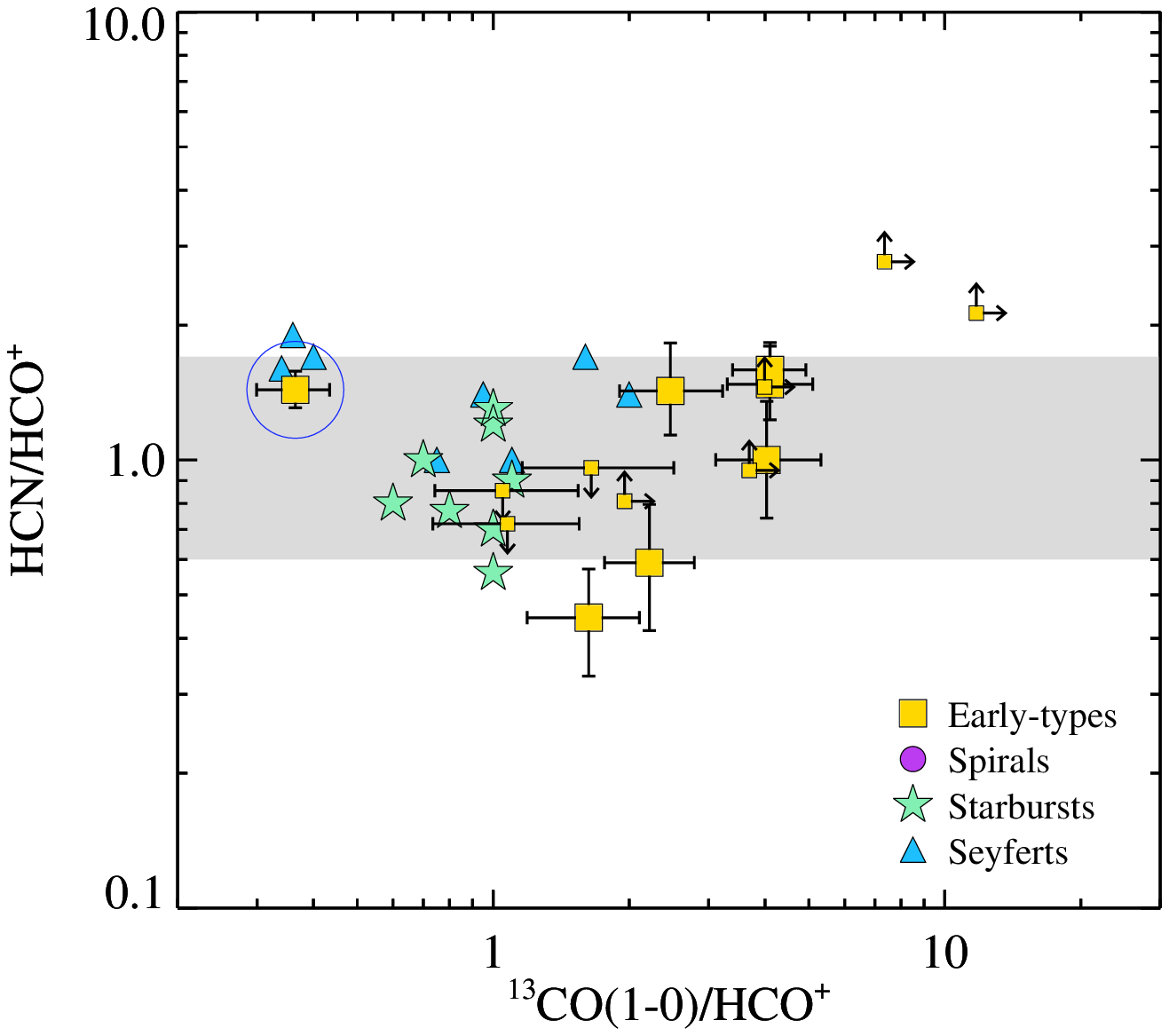}

\caption{Molecular line ratio diagrams. Early-type galaxies are represented by yellow squares with error bars or upper or lower limits. Spirals, starbursts and Seyfert nuclei are portrayed by purple circles, green stars and blue triangles, respectively \citep{krips10}. The grey band highlights the region occupied by spirals from \citet{paglione01} in the top plot, spirals from \citet{gao04a} in the middle plot, and GMCs in M31 \citep{brouillet05} in the bottom plot. The band reflects the fact that the ratio along one axis is not yet well constrained by literature data. NGC~1266, a sample galaxy with a molecular outflow, is identified by blue circles. Cen A is represented by yellow squares without error bars.}
\label{fig:ratioplots}
\end{center}
\end{figure}

\section{Results}

\subsection{Consistency of line profiles}

Assuming that the intrinsic velocity dispersion of the molecular gas is low ($<$ 30 \kms, the channel width), the spectral profile of each molecular line is governed by the velocity field of the galaxy coupled with the spatial distribution of the gas emitting in that line. Changes in spectral shape from line to line can indicate different molecular gas properties at different locations within a galaxy. We thus compare the line profiles of each galaxy, testing for consistency as documented in Appendix~B. Briefly, we first determine whether a Gaussian or a double-peaked profile is the best fit to the \two line. Then, using this best fit functional form, we determine whether the shape parameters (leaving amplitude free) for all other lines are consistent with those derived for the \two line. 

Three galaxies, NGC~1266, NGC~3607 and NGC~4526, have line profiles inconsistent with \two for all or almost all of their other lines. For NGC~1266 (best fit by a Gaussian), the FWHM is significantly greater in its \twt, \hcno and \hcopo lines and is significantly narrower in both \thc lines (further discussed in Section~4.3). \citet{alatalo11} find that NGC~1266's \two and \twt line profiles  are best fit by two nested Gaussians (with a shared centre). We thus perform such a nested two-Gaussian fit for the other molecular lines of NGC~1266, but find that the two-Gaussian fit does not result in a significantly lower $\chi^{2}$ value in these other lines.  For NGC~3607 and NGC~4526, the distinct changes in shape are consistent with pointing offsets which preferentially miss one side of their double-peak profiles.

Out of all the galaxies, NGC~5866 has the most interesting diversity of line profiles. Despite an asymmetry likely due to pointing in both \twc lines, it is clearly double-peaked in both \twc lines, \hcno and \hcopo, while both \thc lines show a clear single central peak. In this galaxy, the molecular gas must have different properties in the centre and disc. 

The line most frequently found to be inconsistent with \two is the \twt line.  These two lines have the highest signal-to-noise ratio and thus their parameters are best constrained, making it harder for them to be formally consistent.  The different beam sizes (factor 2) means a smaller region is traced in \twt providing one reason for potential line-profile differences along with gas property variations and, additionally, the \twt is very sensitive to pointing offsets, as clearly seen in NGC~3607, NGC~3665, NGC~4526 and NGC~5866. 

At some level, we expect the intrinsic line profiles in the various lines to all be different, due to different local conditions in the molecular gas. Higher signal-to-noise data should show such varying profiles and allow variations in the physical properties of the molecular gas to be studied. However, the relatively low signal-to-noise ratio of our observations does not allow for such detailed work and we thus only consider integrated intensity ratios for the remainder of this paper. 

\subsection{Line ratio diagnostics}

Molecular line ratios are a first step to diagnose the state of the molecular gas within galaxies. Here we compare the line ratios of our early-type galaxies with those of spirals, Seyfert nuclei and starbursts. We also test if the molecular line ratios correlate with other ISM  and stellar properties of the galaxies. 

We compute the integrated line ratios after using interferometric \two data (Alatalo et al., in prep.) to  estimate beam correction factors for each galaxy. Appendix~B tabulates these values and describes the process. This step is necessary, as the transitions we compare are at different frequencies, thus measured over different beam sizes.  Corrections from HCN or \hcop to \two are up to 60\% in the case of the most centrally-concentrated sources, but the corrections for most sources range between 30-50\%. All corrections are under 10\% between the \twc and \thc lines and no correction is required between \hcno and \hcopo as they are very close in frequency. 

Errors on the integrated intensity ratios are calculated using a very simple Monte Carlo simulation of errors through division. We include the measurement errors of Table~\ref{tab:ii} and, for ratios of lines not measured simultaneously (i.e. all except HCN/\hcop), a 10\% systematic error contribution. This simulation is necessary because the non-linear operation of division produces a strongly biased (non-Gaussian) error distribution, which is not well approximated by the standard error propagation formula. The bias is particularly noticeable here because our measurements are small compared to their errors (i.e. low signal-to-noise; large relative errors push the first-order Taylor expansion of the standard error propagation formula beyond its region of validity). Using 1000 samples from a Gaussian distribution around each integrated intensity and dividing, we then calculate upper and lower errors by sorting the obtained values and identifying the values that exclude the bottom and top 15.9\% (leaving the central 68.2\% range of values, thus akin to 1$\sigma$ errors on a Gaussian distribution).

\subsubsection{ \twc/\thc}
The sample galaxies vary by a factor of 10 in both \two/\tho (hereafter \Rone) and \twt/\tht (hereafter \Rtwo; see Table~\ref{tab:ratios}) and these two ratios are well correlated (see the top panel of Fig.~\ref{fig:ratioplots}). However, about half the sample galaxies have significantly larger \Rone than \Rtwo ratios, as can be seen comparing the galaxies to the solid 1:1 line in Fig.~\ref{fig:ratioplots} (error bars do not intersect the line). The galaxy area probed by the (1-0) transition is double in diameter to that of the (2-1) transition, so the higher \Rone ratios may indicate that the more extended gas traced by the (1-0) transition is more optically thin in some early-type galaxies. However, the difference in the (1-0) and (2-1) ratios may instead be dominated by the global temperature and density of the molecular gas. Spatially-resolved studies will have to determine what drives the higher \Rone compared to \Rtwo ratios.


The highest \Rone and \Rtwo ratios for the early-type sample are similar to those of the most extreme starbursts, while the lowest ratios have never before been seen in integrated galaxy measures, matching the very lowest values ever measured in galaxy discs \citep{tan11}. As mentioned above, spirals have radial gradients in \Rone, the value generally decreasing with radius in the disc \citep{paglione01, tan11}, although it can have local maxima within spiral arms or intense star-forming regions \citep{schinnerer10,tan11}. In general, high \Rone ratios seem to signal more active star formation by tracing the diffuse gas created by feedback processes. The majority of early-type galaxies have \Rone ratios similar to those of spirals (5-20), with NGC 1222, NGC 1266 and UGC09519 being exceptions at the high end and NGC 4526, NGC 4459 and NGC 3665 being exceptions at the low end. 

\begin{figure*}
\label{fig:sampleplots}
\begin{center}
\subfloat[]{
\label{fig:sampleplots:a}
\includegraphics[height=6.9cm]{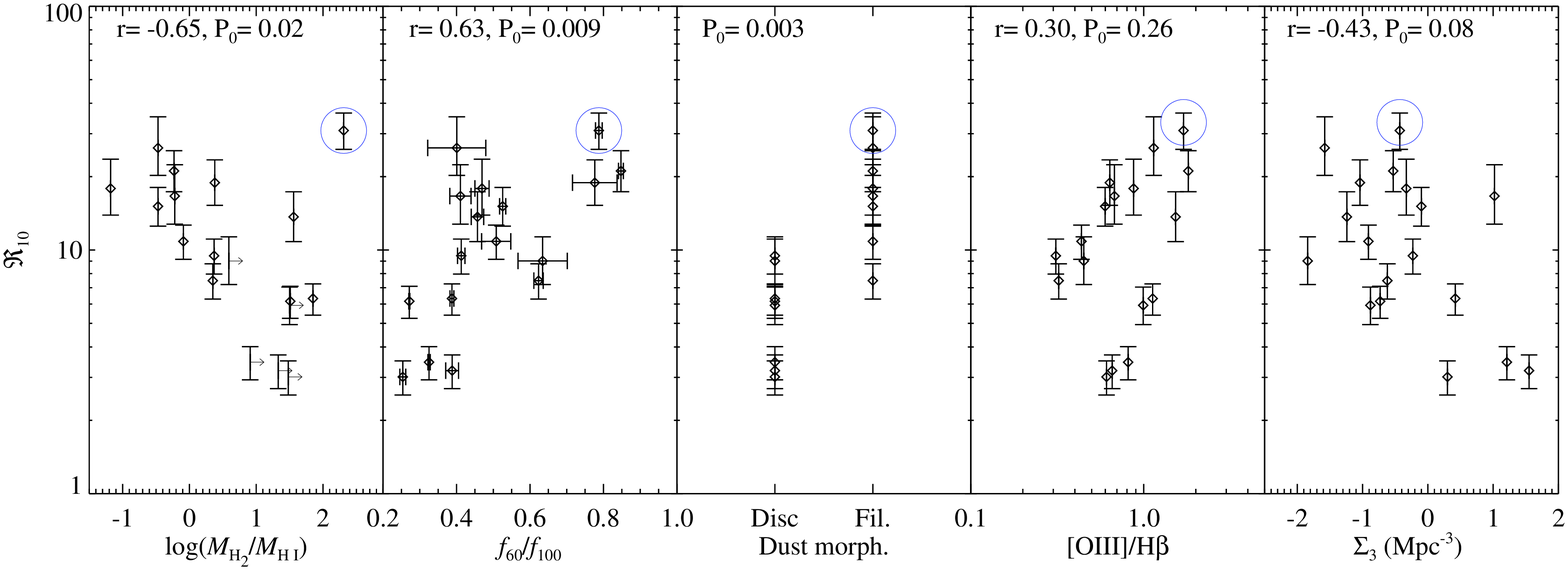}
}\\
\vspace{-5mm}
\subfloat[]{
\label{fig:sampleplots:b}
\includegraphics[height=6.9cm]{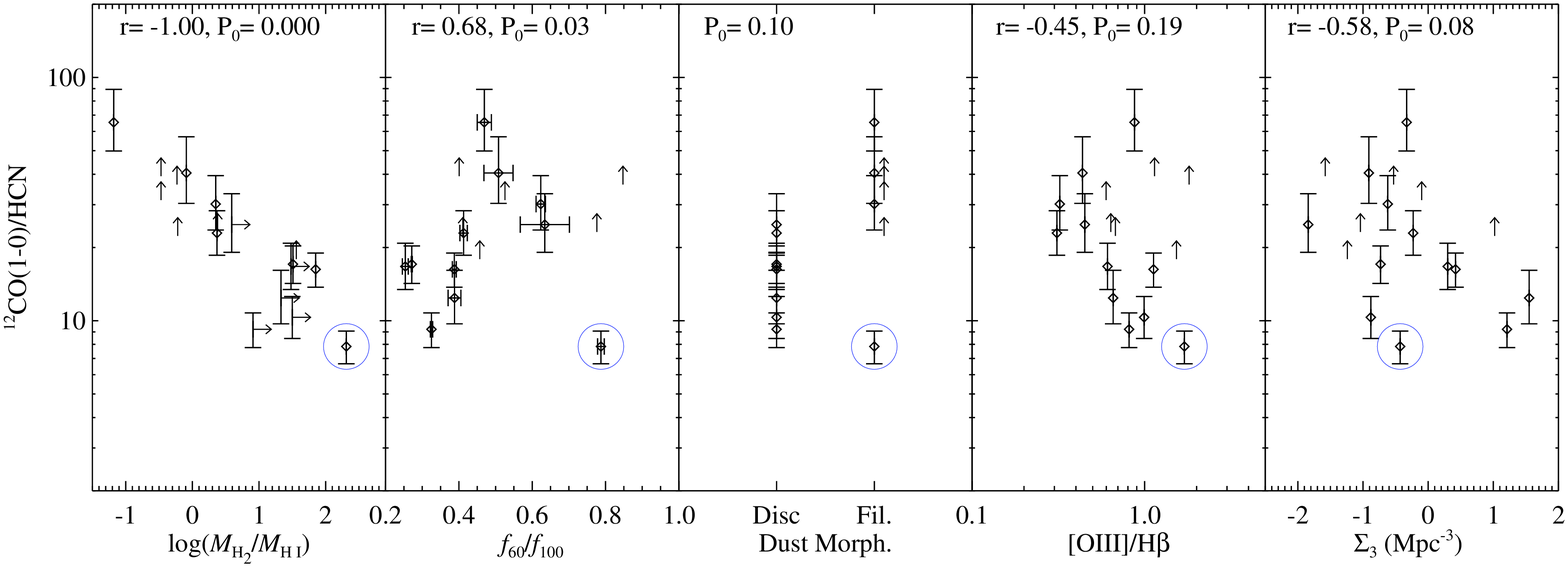}
}\\
\vspace{-5mm}
\subfloat[]{
\label{fig:sampleplots:c}
\includegraphics[height=6.9cm]{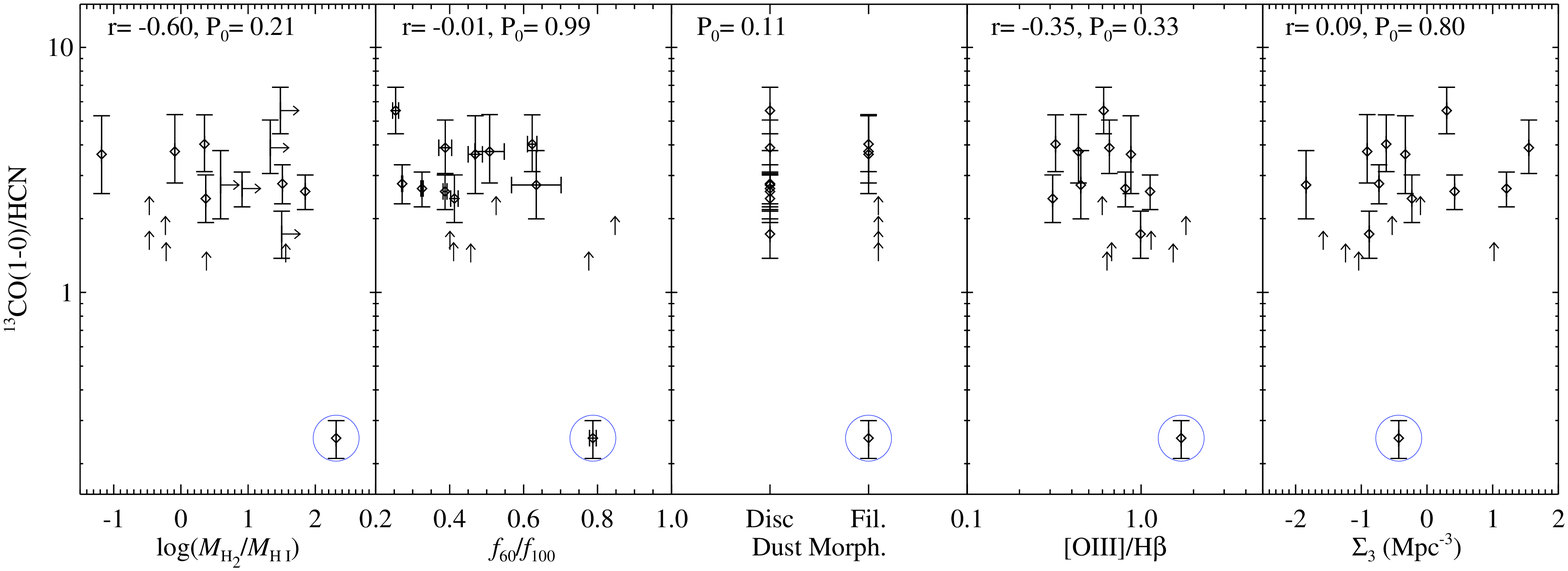}
}
\caption{Molecular line ratios against ISM properties and environment (from left to right): (1) molecular-to-atomic gas ratio, (2) $f_{60}/f_{100}$ ratio of IRAS fluxes (a dust temperature proxy), (3) dust morphology (disc or filamentary; from Paper II), (4) median \oiii/\hbeta emission line ratio (a measure of the ionised gas excitation; from Sarzi et al., in prep.), and (5) $\rho_{10}$ (a measure of the local galaxy density; from Paper VII). From top to bottom, the molecular ratios are (a) \Rone, (b) \two/HCN  and (c) \tho/HCN.  NGC~1266 is highlighted with a circle. The Spearman rank correlation coefficient (`r') and the probability that the null hypothesis (`P$_{0}$', no correlation) is true are given at the top of each plot. }
\end{center}
\end{figure*}

\begin{figure*}
\label{fig:lsplots}
\begin{center}
\subfloat[]{
\label{fig:lsplots:a}
\includegraphics[height=7cm]{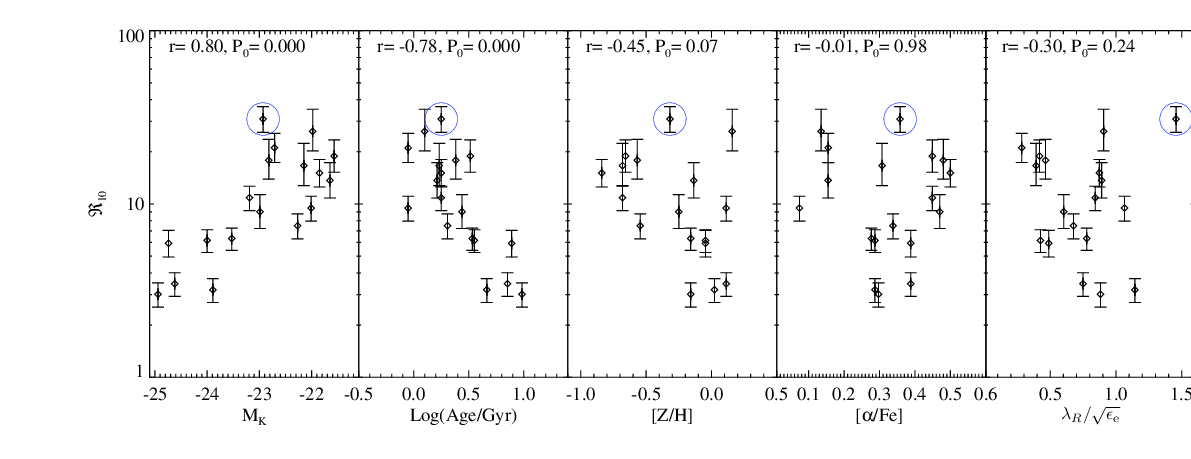}
}\\
\vspace{-5mm}
\subfloat[]{
\label{fig:lsplots:b}
\includegraphics[height=7cm]{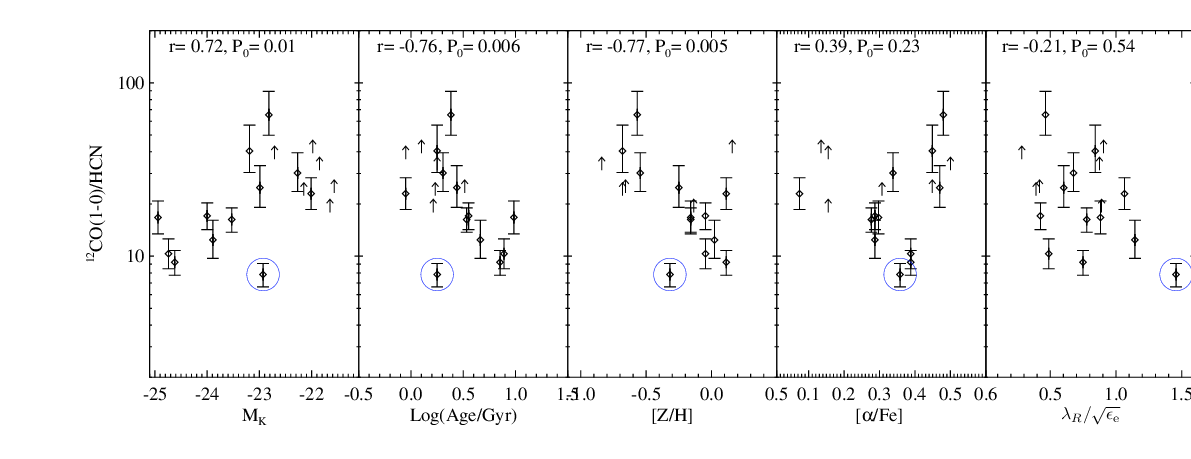}
}\\
\vspace{-5mm}
\subfloat[]{
\label{fig:lsplots:c}
\includegraphics[height=7cm]{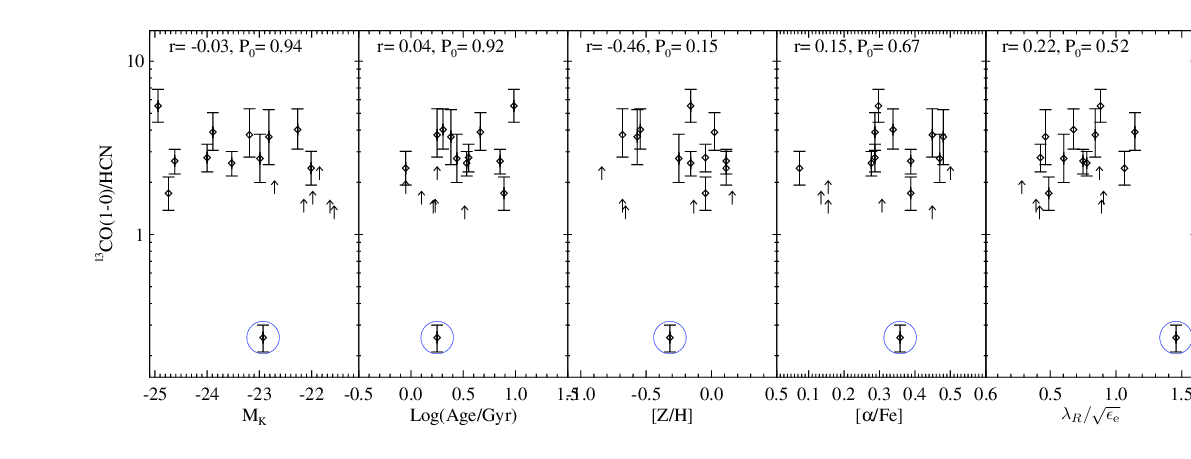}
}
\caption{Molecular line ratios against stellar properties (from left to right): (1) absolute $K$-band magnitude, SSP best-fit (2) age, (3) metallicity, (4) alpha-element abundance (McDermid et al., in prep.) and (5) specific stellar angular momentum normalized by ellipticity (Paper II).  From top to bottom, the molecular ratios are (a) \Rone, (b) \two/HCN  and (c) \tho/HCN. NGC~1266 is highlighted with a circle.The Spearman rank correlation coefficient (`r') and the probability that the null hypothesis (`P$_{0}$', no correlation) is true are given at the top of each plot.}
\end{center}
\end{figure*}

Figures~\ref{fig:sampleplots:a} and \ref{fig:lsplots:a} show how the \Rone ratio varies with several galaxy properties, Figure~\ref{fig:sampleplots:a} focusing on ISM properties and environment and Figure~\ref{fig:lsplots:a} focusing on stellar properties.  The galaxy with a significant molecular outflow, NGC~1266, is identified in each plot with a circle and will be discussed separately. At the top of each plot, the Spearman rank correlation coefficient and the probability that the null hypothesis (no correlation) is true are given. Low probabilities for the null hypothesis (we take `low' to be less than 0.03) suggest the data are correlated. The dust morphology has two discrete categories (disc and filament), so the null hypothesis is that the sample means are the same. We evaluate this probability by Welch's t-test (similar to the Student t-test but not assuming that the sample variances are the same). 

By the above measures, \Rone is correlated with the molecular-to-atomic gas ratio (atomic gas measured by \hi emission), the dust temperature, M$_{{K}}$ and the single stellar population (SSP) derived age. The \Rone ratios of the galaxies with dust discs are also significantly different than those of galaxies with filamentary dust. On the other hand, no significant correlation is found with the [OIII]/\hbeta emission line ratio (a measure of the ionised gas excitation), local environment ($\Sigma_{3}$), SSP-derived metallicity or alpha-element enhancement, or specific stellar angular momentum ($\lambda_{\mathrm{R}}/\sqrt{\epsilon_{\mathrm e}}$).

Of course, not all of these parameters are independent. Stellar luminosity (or mass) is known to positively correlate with the SSP-derived age, metallicity and alpha-element enhancement \citep[][McDermid et al., in prep.]{trager00,thomas05}. The bright-CO selection criterion results in a relation between stellar mass and gas fraction such that galaxies with higher stellar mass have low gas fraction and vice versa. The galaxies with higher gas fractions tend to be those that are interacting, with lower molecular-to-atomic gas ratios, filamentary dust morphologies and higher dust temperatures, while the low gas fraction galaxies are more molecular, with settled and colder dust discs (see Table~\ref{tab:sample}). We note that low mass galaxies with low gas fractions (and settled dust morphologies and low dust temperatures) do exist among the 56 CO-detected \atlas galaxies. Observing such galaxies will help separate some of these currently degenerate parameters. To proceed with the current sample, we draw from the most plausible of relations, supported by previous work on late-type galaxies.

 It has long been known that \Rone directly correlates with the dust temperature in spiral and starburst galaxies \citep[e.g.][]{young86,aalto95}. The explanation proposed  is that hot dust temperatures signal more efficient star formation (more UV photons impinging on a given quantity of dust), that in turn may produce optically thin molecular outflows or a hot, diffuse component of molecular gas through radiative feedback. In either case, higher \Rone ratios result. 
 
All of the sample galaxies with high \Rone ratios have filamentary dust structure, indicating an unsettled ISM. Signs of interaction are also often visible, either in stellar light (Paper II, \citealp{duc11}, hereafter Paper IX) or in tails of atomic gas  (Serra et al., in prep.). However, unlike the direct relation between \Rone and $f_{60}/f_{100}$ seen in \citet{young86} and \citet{aalto95}, a number of the early-type galaxies with high \Rone do not have particularly warm dust (IC~1024, NGC~4694, NGC~7465 and UGC~09519). These galaxies are interacting or accreting, but do not (currently) have starburst or nuclear activity substantially heating their dust.  For these galaxies, the low optical depth in \two must be attributed to unsettled gas as opposed to feedback from intense star formation. Indeed, interacting galaxies have been observed to have increased gas velocity dispersions \citep{irwin94,elmegreen95}, which would allow for a decreased optical depth in \two. 

Thus, the \Rone ratio seems driven to higher values by gas-rich interactions (major mergers, minor mergers, or tidal gas capture) and/or feedback from starburst activity. The gas-rich interactions tend to have large amounts of atomic gas involved, thus explaining the correlation with the molecular-to-atomic gas ratio. Dust temperatures of these galaxies are middle to high, never very low. The strong apparent correlation with $K$-band luminosity may be due to our sample selection, since we have no high mass, high gas-fraction (generally interacting) galaxies or low mass, low gas-fraction (generally settled) galaxies. Similarly, the strong trend with SSP age may be tied to the biased gas fraction because SSP age is very sensitive to the fraction of young to old stars (which is presumably higher in high gas fraction galaxies). As mentioned above, including low mass, low gas-fraction galaxies in future work will allow us to see if either galaxy mass or SSP age affect the \Rone line ratio. 

 
 In the fourth panel of Fig.~\ref{fig:sampleplots:a}, the eye is drawn to a correlation between \oiii/\hbeta, although the Spearman correlation coefficient reveals this is not particularly likely. Indeed, it is hard to state a connection: high \Rone galaxies shy away from the lowest values of \oiii/\hbeta while low \Rone galaxies have middle valued-ratios. Coupled with another optical emission line ratio such as \nii/\halpha or \sii/\halpha, \oiii/\hbeta is frequently used to distinguish HII-region photoionisation from shock ionisation or from a power-law continuum source such as an AGN. Unfortunately, without the second line ratio, the values observed for our galaxies are degenerate. Thus the lack of a clear correlation may be due to various ionisation sources which are differently (or not) related to molecular gas properties. 
 
 Some relation also looks possible for \Rone and $\Sigma_{3}$, mostly driven by galaxies at high local environmental densities having low \Rone values. These are all galaxies that are in the Virgo cluster or the centre of their group. It is possible that in these galaxies, the molecular gas is additionally pressure confined by the hot intra-group or intra-cluster medium and so has such high optical depth. But more galaxies are needed to rule in such a scenario. 
 
We have suggested that either stellar feedback and/or ongoing accretion of cold gas leads to more optically thin molecular gas and higher \Rone ratios. However, chemical effects may also be able to explain such relations. $^{13}$C is a secondary product of nucleosynthesis and should build up in abundance relative to $^{12}$C as a stellar population ages and returns material to the interstellar medium. Interactions may bring less enriched material into galaxies, with presumably less $^{13}$C, explaining the higher \Rone ratio seen in these systems. Chemical fractionation due to isotopic ion exchange towards \thc and away from \twc may also cause the trend with $f_{60}/f_{100}$; at lower temperatures the lower-energy \thc will be more strongly preferred. If the majority of molecular gas is at low temperatures in the more settled galaxies, then this option could also be responsible for the observed trends. Ratios involving HCN and \hcop help disentangle these options in the next sections.
 
We note that three of the early-type galaxies have extremely low \Rone and \Rtwo ratios, at the limit seen in spiral galaxy discs: NGC~3665, NGC~4459 and NGC~4526. A \Rone minimum value of 3.3 was reported in \citet{tan11}, but spiral discs generally have \Rone values above 5 \citep{young86, paglione01, schinnerer10}. The values observed here indicate that the gas in these three galaxies is either less affected by star formation feedback (remaining very optically thick and/or fractionated towards \thc) or is more enriched with $^{13}$C. 

\subsubsection{HCN}
For HCN, we consider ratios with both \two and \tho (see Table~\ref{tab:ratios}).  HCN/\two varies by a factor of about 7, from 0.015 to 0.108, while HCN/\tho varies by only  a factor $\approx$3 from 0.18 to 0.58, if NGC~1266 is excluded (because of its unique properties associated with its molecular outflow). Our early-type galaxies mostly lie within the HCN/\two range of normal spirals indicated by the grey band in Fig.~\ref{fig:ratioplots}, based on the galaxies of \citet{gao04a} with far-infrared luminosity $L_{\mathrm{FIR}} < 10^{11}$ L$_{\sun}$ (thus below the LIRG luminosity threshold). The three galaxies above the grey band are NGC~3607, NGC~4459 and NGC~4526. These are three galaxies that were also low in \Rone and \Rtwo, signaling that it may be their weakness in \twc that is the shared trait between the two ratios. NGC~7465 lies below the grey band; it is a galaxy with high \Rone and \Rtwo.  Few spirals have measured HCN/\tho ratios, but the early-types definitely have HCN/\tho ratios lower than starbursts (even those with high starburst-like \Rone ratios), while they overlap with approximately half of the Seyfert nuclei.  

We plot the inverse of these ratios (thus \two/HCN and \tho/HCN) against the same galactic properties as before to make plots with the same orientation as for \Rone (Figs.~\ref{fig:sampleplots:b},c and \ref{fig:lsplots:b},c). Again ignoring NGC~1266, very similar trends to those of \Rone are  seen with \two/HCN, although the lower number of galaxies (only 12 detected in HCN) makes it easier for the null hypothesis of no correlation to be accepted. Correlations with the molecular-to-atomic gas ratio, dust temperature, SSP age and M$_{{K}}$ are still significant. A correlation with stellar metallicity is also shown to be likely, but would probably not be if the one lower limit at high Z were included. For the \tho/HCN ratio, no correlation is seen with any galaxy parameter.

In the \two/HCN ratio, there is no difference in the carbon isotope, so the continued presence of most of the trends observed with \Rone rules out enhanced $^{13}$C abundance as their cause. If ion-exchange reactions leading to fractionation were responsible for the trends with \two/HCN, inverse relations would be expected for \tho/HCN. As this is not seen, fractionation is also unlikely to explain the observed trends. With both abundance variations and ion-exchange reactions unable to drive the observed trends, a varied average optical depth of \two is left as the most likely cause.

\begin{figure}
\begin{center}
\includegraphics[width=7.5cm]{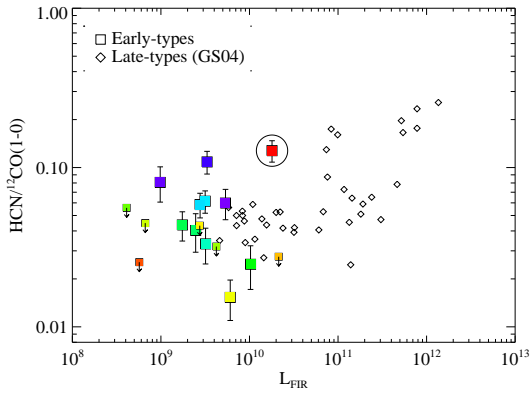}

\caption{HCN/\two versus FIR luminosity. Early-type galaxies are shown as colored squares with the color reflecting their \Rone ratio (blue to red depicts low to high ratios). Late-type galaxies from the \citet{gao04b} sample are shown as diamonds. NGC~1266 is highlighted by a surrounding circle.}
\label{fig:gao}
\end{center}

\end{figure}

HCN/\two is frequently used as a dense gas tracer in star-forming galaxies. \citet{gao04b} find an increased HCN/\two ratio for LIRG and ULIRG galaxies and suggest that gas traced by HCN is more closely tied to star formation than that traced by \two. In Fig.~\ref{fig:gao}, we place our early-type galaxies on a HCN/\two versus $L_{\mathrm{FIR}}$ plot along with the late-type galaxies from \citet{gao04b}. The early-type galaxies at low $L_{\mathrm{FIR}}$ do not behave as nicely as the late-types, instead they exhibit a wide range of HCN/\two ratios. However, much of this may be due to the optical depth effects in \two described above. We colour-code the early-type galaxies by their \Rone ratio, and as expected if the optical depth of \two plays a role, galaxies with low \Rone ratios are seen at high HCN/\two and vice versa (a blue colour indicates high HCN/\two). Thus HCN/\two is not a good tracer of dense gas fraction in early-type galaxies, HCN/\tho should be better. Additionally, we see that NGC~1266 has a high HCN/\two ratio despite its high \Rone ratio. It thus exhibits a dense gas ratio like a LIRG or ULIRG regardless of optical depth considerations. Thus high dense gas fractions are not limited to galaxies that are extremely bright in the FIR. 

If we instead use HCN/\tho as a dense gas tracer, we find that that there is only about a factor 3 variation in this ratio (ignoring NCC~1266) among the early-type galaxies.  Unfortunately, few spiral galaxies have both HCN and \tho measured; only a comparison with a larger spiral sample will really tell whether the early-type galaxies as a population have different dense gas fractions.

\subsubsection{HCO$^{+}$}
\hcopo is detected in 10 of the 18 sample galaxies, including 3 cases where HCN is not detected (NGC~1222, PGC~058114 and UGC~09519). The HCN/\hcop ratios for the detected early-type galaxies range from 0.44 (and possibly less) to $>2.77$. Ratios of 0.66 to 1.7 are seen in M31 giant molecular clouds \citep[GMCs][]{brouillet05}; this region is shaded in grey in the bottom panel of Fig.~\ref{fig:ratioplots}. Most early-type galaxies are also found within this range, as are most starbursts and Seyferts, although Seyfert nuclei appear biased towards higher and spiral galaxies towards lower HCN/\hcop ratios \citep[e.g.][]{krips08}. As for the \tho/\hcop ratio, the early-type galaxies have larger ratios than the starbursts (and many of the Seyferts), analogous to the smaller HCN/\tho ratios observed. Thus both ratios agree that early-type galaxies have lower dense gas fractions than starburst (and many Seyfert) galaxies, at least if the latter systems are not dominated by chemical enhancement of the HCN and \hcop.

Curiously, two of our galaxies (NGC~3665 and NGC~4526) have quite high HCN/\hcopo ratios, 2 or higher based on our \hcop non-detections. These are two of the three galaxies with extremely low \Rone values. The third very low \Rone galaxy (NGC~4459) is also a non-detection in \hcop; its HCN/\hcop must be above $1.4$. None of these three galaxies has a clear Seyfert nucleus based on optical emission line ratio classification \citep{ho97a} or a powerful X-ray nucleus, although one (NGC~3665) has conspicuous radio jets that may signal an X-ray bright AGN hidden within. Still, an AGN driving XDR chemistry cannot be the common link between these three galaxies. Instead, the possible low efficiency of star formation in these galaxies may support a lower ionisation fraction and thus a smaller abundance of \hcop relative to HCN. Information from other molecular gas tracers, such as HNC, will help diagnose if this is the case.

\subsection{Molecular gas classes}

As mentioned above, early-type galaxies exhibit a variety of molecular line ratios. Here we draw attention to two basic classes: galaxies with settled dust and gas discs and those with unsettled dust and gas. 

Galaxies with settled discs include NGC~3032, NGC~3607, NGC~3665, NGC~4459, NGC~4526, NGC~4710, NGC~5866 and NGC~6014. Most also have clearly double-peaked spectra, indicating that the molecular gas distribution extends beyond the turnover of the galaxy's rotation curve (Paper V), and all have more molecular than atomic gas. These galaxies have low \Rone and \Rtwo values, in three cases strikingly low (below 3). They also have high HCN/\two ratios, consistent with the idea that they are very optically thick in \two. Many also have high ($>$ 1.4) HCN/\hcopo ratios. 

Within this set of settled gas discs, we find that those with the greatest $K$-band luminosities have the lowest \Rone, \Rtwo and \two/HCN ratios. If these galaxies also have the least efficient star formation (as predicted by morphological quenching for galaxies with high bulge to disk ratios) then the very optically thick molecular gas in these galaxies might result from reduced mechanical and/or radiative feedback accompanying the less efficient star formation. Another option is that they are generally found in environments where hot gas may confine their molecular gas to a thin (and thus more optically thick) layer. 

The second class of galaxies we draw attention to are those with unsettled dust (Paper II), signaling either ongoing gas accretion or a merger: IC~676, IC~1024, NGC~1222, NGC~2764, NGC~4694, NGC~7465 and UGC~09519. Recent work suggests that minor mergers are likely to supply the cold gas to many of the currently star-forming early-type galaxies \citep{kaviraj09}, so such signs of accretion are not unexpected. These galaxies generally have more atomic than molecular gas and the atomic gas distributions frequently, although not always, show signs of interaction \citep{serra11}. PGC~058114 does not have optical photometry for dust classification, but generally seems to fit in this class. These galaxies all have high \Rone ratios and low HCN/\two ratios (see Fig.~4). Increased turbulence from the interaction, diffuse molecular gas within the significant atomic phase and/or significant radiative or mechanical feedback from a related starburst creates a lower optical depth in these galaxies. This group has a wide spread in dust temperatures.

 NGC~4150 may bridge the two classes; recent HST WFC3 imaging reveals a dust disc with spiral structure and stellar population analysis suggests a recent ($\approx 1$~Gyr) minor merger \citep{crockett11}. \hi observations show a low-column density cloud of atomic hydrogen, possibly leftover from this minor merger \citep{morganti06}. However the overall molecular-to-atomic gas fraction is quite high, making NGC~4150 more similar to members of the settled disc group.  Its molecular line ratios lie in the middle of the ranges we observe. In general, there is likely to be a continuum from highly disordered to settled ISM. Many of the interacting galaxies may have or may be developing a disc, and some of the galaxies with discs may not be entirely settled yet. 

As mentioned previously, NGC~1266 stands apart in many ways from the other galaxies. It has a massive molecular outflow, discovered during the IRAM~30m survey of the \atlas galaxies \citep{alatalo11}. Here we note its extremely high HCN/\tho ratio of 2.7,  higher than most starbursts and Seyfert nuclei found in the literature. The interpretation of this ratio as an extremely high dense gas fraction fits with the very high surface density implied by the interferometric \twc observations. But an increased ionisation rate due to X-rays, cosmic rays, or very intense FUV emission could also boost the HCN abundance. The clear line profile mismatches suggest spatial variations in the molecular properties or multiple molecular components.  The \twt, \hcno and \hcopo lines were all fit with wider Gaussians (FWHM = 154$^{+1.4}_{-1.4}$, 152$^{+15}_{-13}$ and 176$^{+26}_{-21}$ km s$^{-1}$, respectively) than the \two line (FWHM =  132$^{+3}_{-3}$ km s$^{-1}$), suggesting that they are more prevalent with respect to \two in the molecular outflow (which reaches higher velocities). On the other hand, the narrow profiles of both the \tho (FWHM = 94$^{+\;9}_{-10}$ \kms) and \tht (FWHM = 106$^{+9}_{-8}$ \kms) lines indicate reduced emission from the \thc lines in the outflow. The outflow must therefore have an even higher \Rone ratio and is likely very optically thin, consistent with the large velocity gradient modeling performed in \citet{alatalo11} using the \two,\twt and $^{12}$CO(3-2) transitions. The lack of strong \thc wings confirms a typical $X_{\mathrm{CO}}$ factor should not be used for the molecular outflow.

\section{Conclusions}

Using the IRAM 30m telescope, we have surveyed the \thc, HCN and \hcop emission of the 18 galaxies with the strongest \twc emission in the \atlas sample of nearby early-type galaxies. We detect all 18 galaxies in both \tho and \tht transitions, 12/18 in \hcno and 10/18 in \hcopo.  

Fitting the \two lines with a double-peak or Gaussian profile and evaluating the reduced $\chi^{2}$ value, we find that one third of the galaxies are best fit by a Gaussian, one third with a clear double-peak profile and the remaining one-third with a very flat-topped `double-peak' profile. Two galaxies (NGC~1266 and NGC~5866) have clearly different profiles in the various molecular lines, indicating spatial variations of the  molecular gas conditions within the galaxies. Discrepant profiles in a few other galaxies are consistent with pointing offsets. We hypothesize that at higher signal-to-noise ratios, the majority of the line profiles would be found to be intrinsically inconsistent due to changing molecular gas conditions. Higher signal-to-noise single-dish data or interferometric maps in the various lines will allow us to determine if this is indeed the case. 

The molecular gas line ratios of early-type galaxies are generally consistent with the ranges observed in spiral galaxies, although we also identify some outliers. In terms of the HCN/\two ratio, often used as a dense gas tracer, we extend the work of \citet{gao04b} to lower FIR luminosities and find an increased scatter in the ratio, compared to late-type galaxies of similar FIR-luminosities. Because of the important effects of the optical depth of \two seen in the \Rone ratio, we suggest that optical depth may be undermining the use of HCN/\two as a dense gas tracer and instead recommend the use of HCN/\tho. Unfortunately, few spirals have literature HCN/\tho ratios. Determining whether the early-types have significantly different dense gas fractions thus requires a larger sample of spirals to be observed in both \hcno and \tho.

We compare the molecular gas ratios \Rone, \two/HCN and \two/\hcop to several ISM and stellar galaxy properties. \Rone is found to correlate with the molecular-to-atomic gas ratio, dust temperature, dust morphology, absolute $K$-band magnitude and SSP age. \two/HCN similarly correlates with the molecular-to-atomic gas ratio, dust temperature, absolute $K$-band magnitude and SSP age. The persistence of most \Rone correlations with \two/HCN rules out abundance variations of \thc as the driving factor of the relations. No correlation with any of these properties is seen in the \tho/HCN ratio, further ruling out a dominant contribution from ion-exchange fractionation towards \thc at low temperatures. 

With these options removed, a change in optical depth of the \two line is the best explanation for the correlations we see. Based on the lower molecular-to-atomic gas ratios, unsettled dust and occasionally high dust temperatures, we find that high \Rone and \two/HCN ratios in early-type galaxies are linked to recent or ongoing interactions and/or starbursts. Both of these processes can naturally account for optically thinner \two line emission, as interactions are known to be more turbulent and star formation feedback can also produce more turbulence or heat up the molecular gas. Early-type galaxies with settled gas and dust discs tend to have less optically thin molecular gas, along with lower dust temperatures and higher molecular to atomic gas mass ratios. The correlations with $K$-band luminosity and SSP age can plausibly be explained by sample selection effects, further work including less massive early-type galaxies with settled gas and dust discs is required before conclusions are reached.  

A set of outliers with low \Rone and \Rtwo, high HCN/\two and only upper limits to their HCN/\hcop ratios are identified. These galaxies may have particularly stable molecular gas which remains very optically thick in \twc  or have their molecular gas confined to a thin layer by hot gas pressure. NGC~1266, a galaxy with a molecular outflow, also stands apart from both spirals and the rest of the early-type sample. Its HCN/\tho and \tho/\hcop ratios suggest that it has a very high dense gas fraction, but its high \Rone and \Rtwo values simultaneously suggest that much of its \twc emission is optically thin. This combination of ratios is seen in some other extremely active starburst or Seyfert galaxies. 

\section*{Acknowledgements}
Based on observations carried out with the IRAM Plateau de Bure
Interferometer. IRAM is supported by INSU/CNRS (France), MPG (Germany)
and IGN (Spain). 

AC would like to thank Suzanne Aalto and Ron Snell for useful discussions during the preparation of this work as well as Leslie Sage and Gary Welch for providing \two and \twt data for two sample galaxies.

This work was supported by the rolling grants `Astrophysics at Oxford' PP/E001114/1 and ST/H002456/1 and visitors grants PPA/V/S/2002/00553, PP/E001564/1 and ST/H504862/1 from the UK Research Councils. EB thanks John Fell OUP Research Fund, ref 092/267. MC acknowledges support from a Royal Society University Research Fellowship. 
RLD acknowledges travel and computer grants from Christ Church, Oxford and support from the Royal Society in the form of a Wolfson Merit Award 502011.K502/jd. RLD also acknowledges the support of the ESO Visitor Programme which funded a 3 month stay in 2010.

FB acknowledges support from the European Research Council through grant ERC-StG-257720.

SK acknowledges support from the the Royal Society Joint Projects Grant JP0869822.

RMcD is supported by the Gemini Observatory, which is operated by the Association of Universities for Research in Astronomy, Inc., on behalf of the international Gemini partnership of Argentina, Australia, Brazil, Canada, Chile, the United Kingdom, and the United States of America.

TN and MBois acknowledge support from the DFG Cluster of Excellence `Origin and Structure of the Universe'.

MS acknowledges support from a STFC Advanced Fellowship ST/F009186/1.

NS and TD acknowledge support from an STFC studentship.

MBois has received, during this research, funding from the European Research Council under the Advanced Grant Program Num 267399-Momentum.

The authors acknowledge financial support from ESO. The research leading to these results has received funding from the European
Community's Seventh Framework Programme (/FP7/2007-2013/) under grant agreement
No 229517.

\bibliographystyle{mn2e}
\bibliography{t_v3.bbl}

\pagebreak
\appendix
\section{Remaining spectra}
The spectra of the remaining 14 galaxies not shown in Fig.~1 are shown here in Fig.~\ref{fig:otherspec}. 

\begin{figure*}
\begin{center}
\begin{tabular}{cccc}
\includegraphics[height=10.5cm]{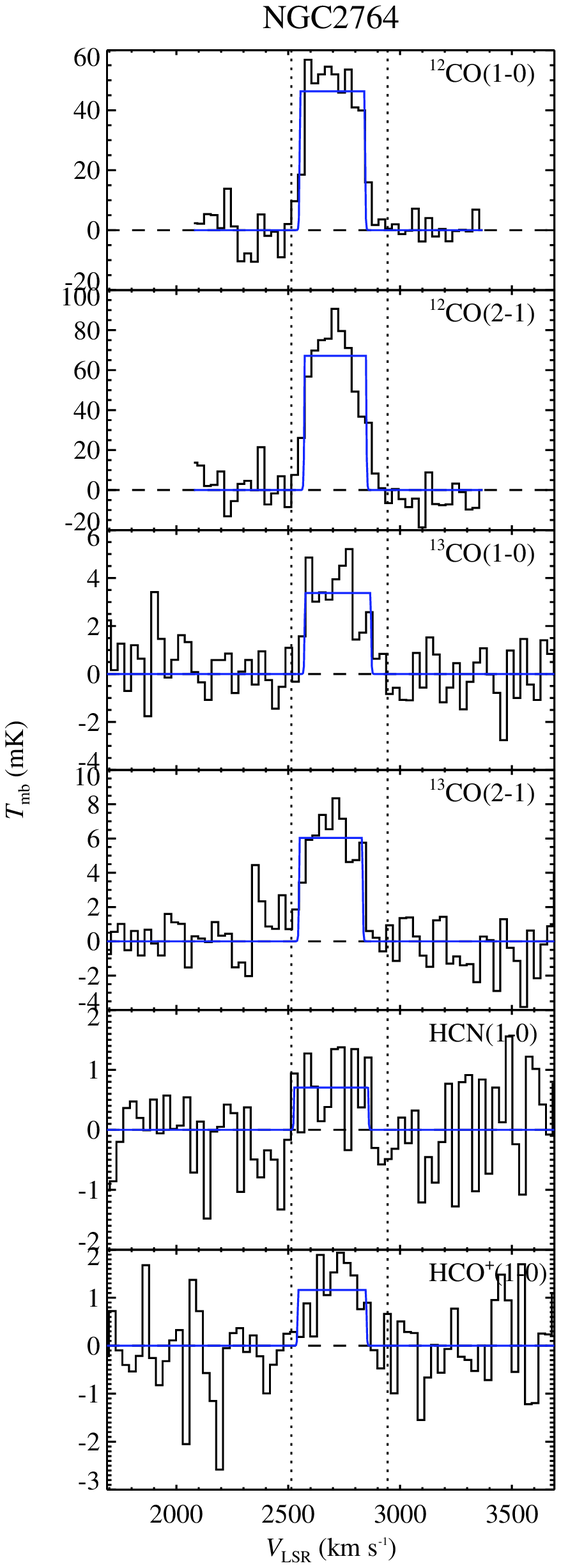} & 
\includegraphics[height=10.5cm]{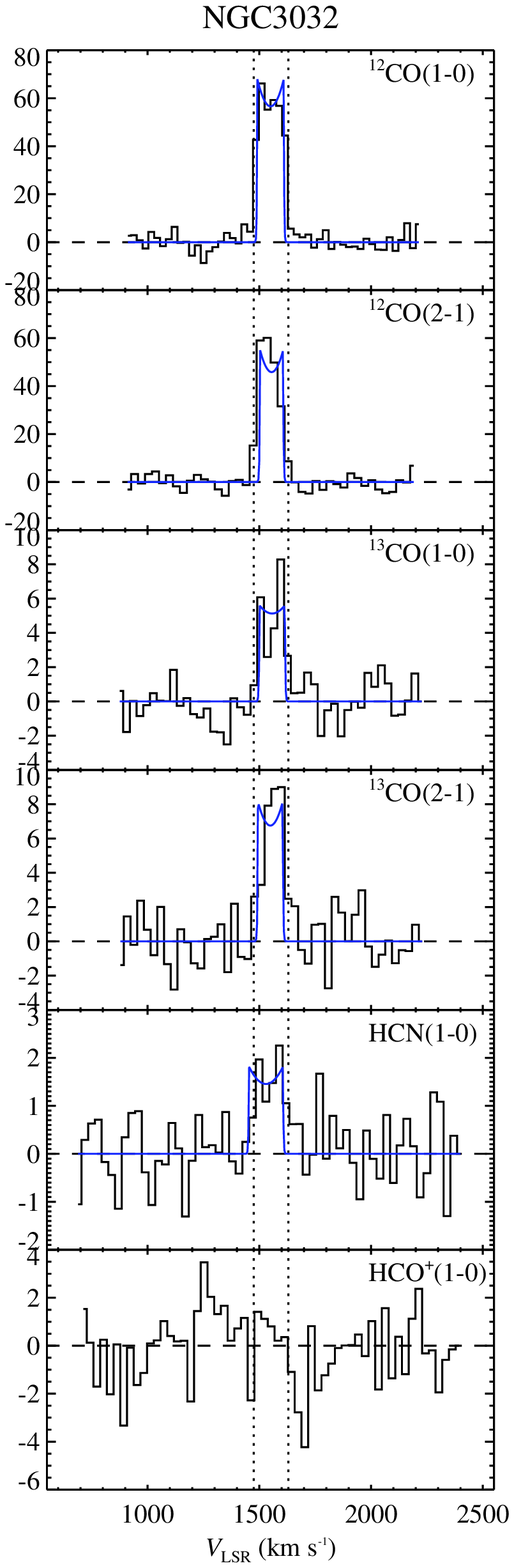} & 
\includegraphics[height=10.5cm]{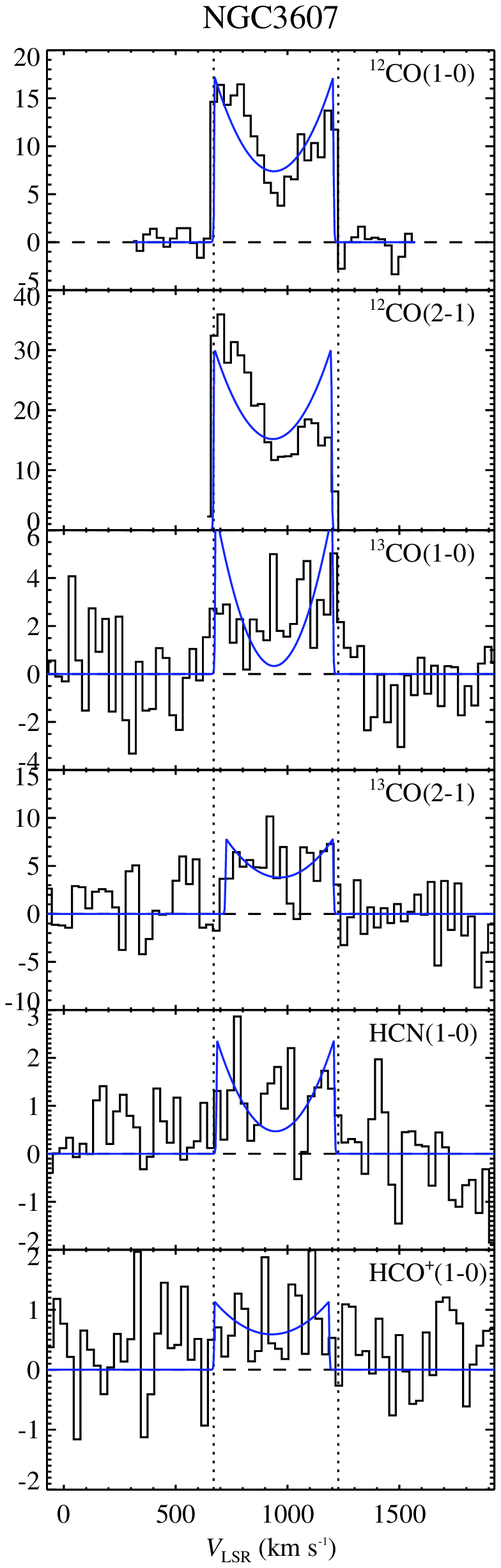} & 
\includegraphics[height=10.5cm]{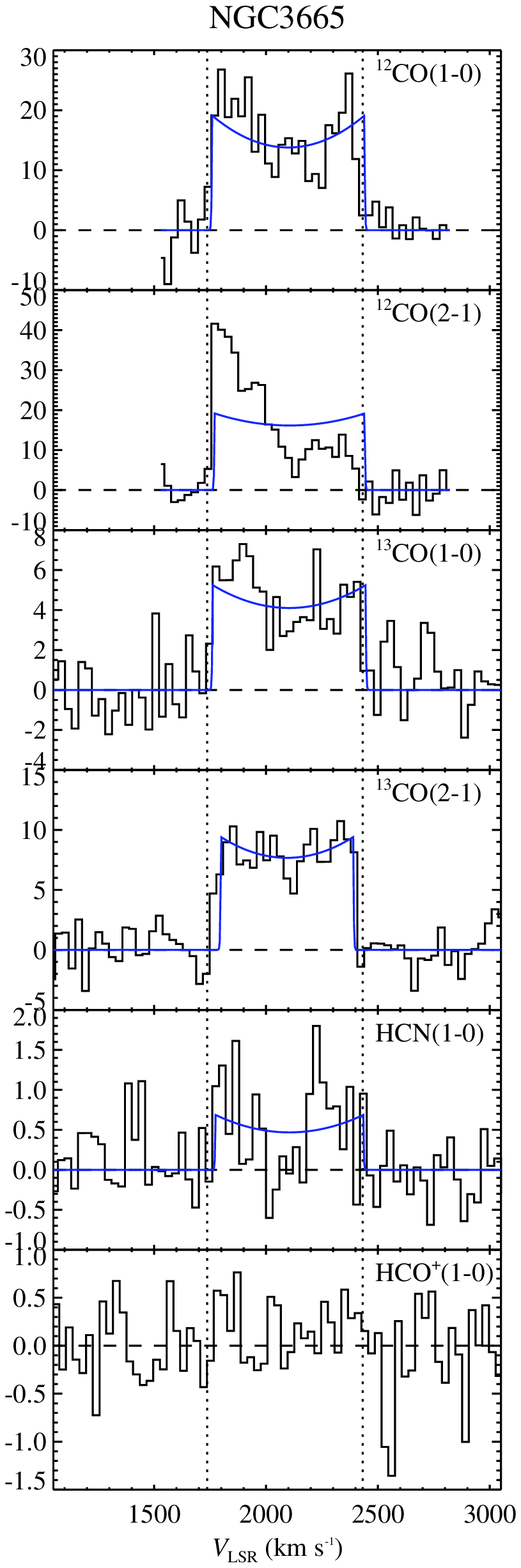}\\
\includegraphics[height=10.5cm]{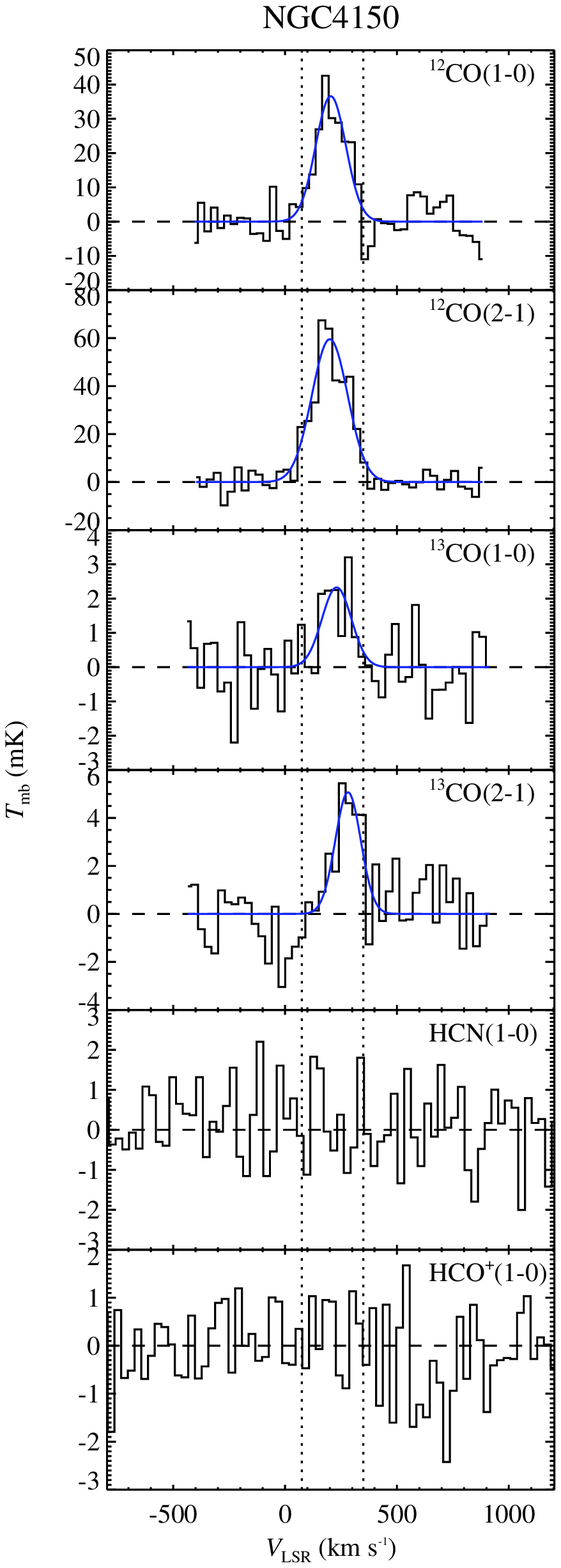} &
\includegraphics[height=10.5cm]{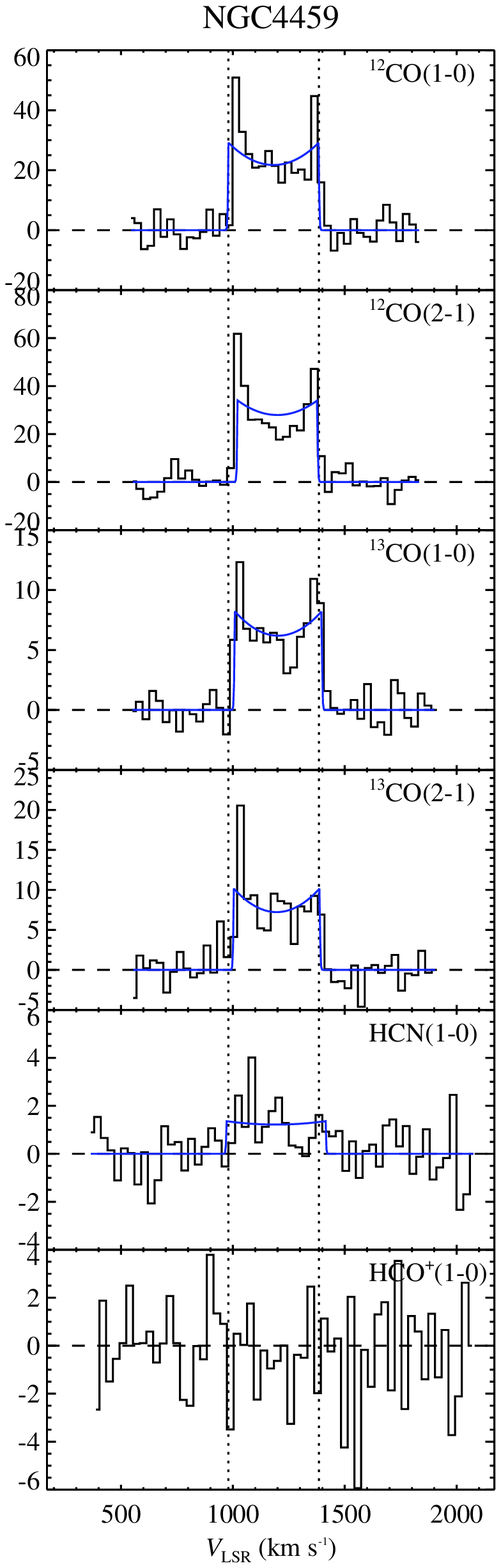} & 
\includegraphics[height=10.5cm]{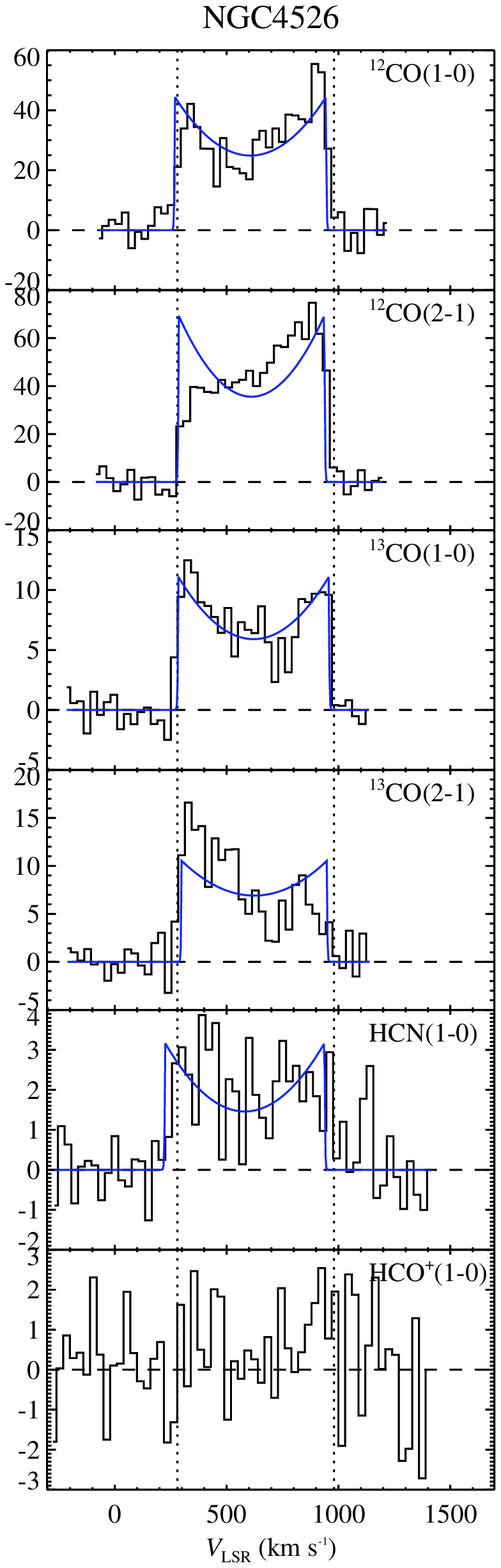} & 
\includegraphics[height=10.5cm]{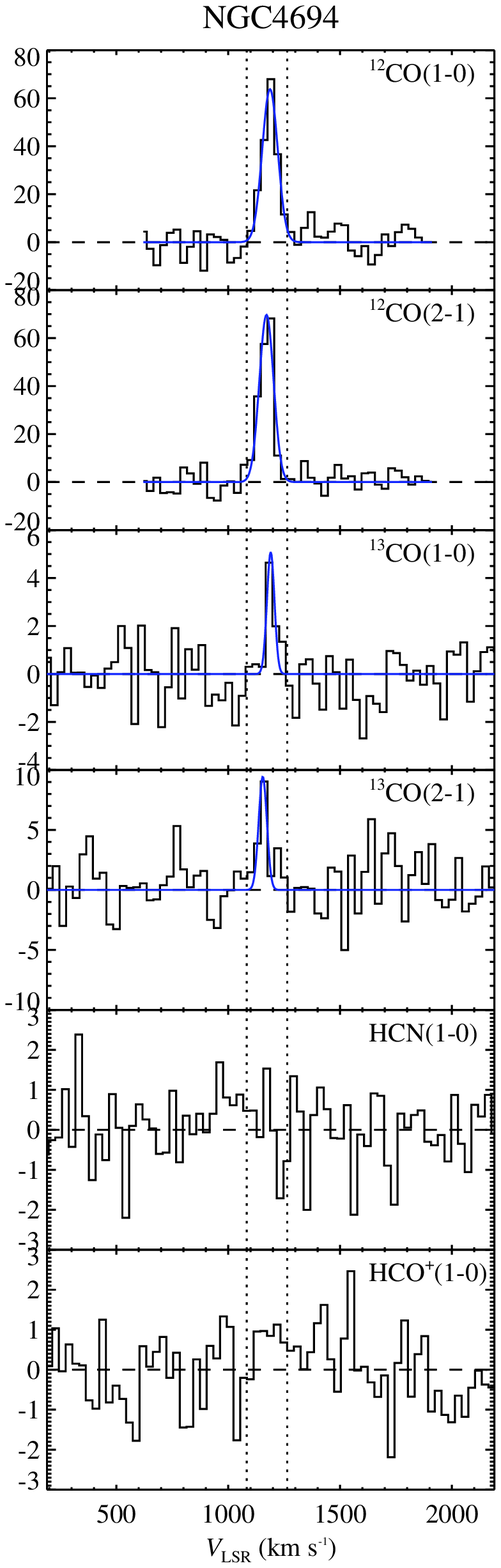}\\
\end{tabular}
\end{center}
\caption{Molecular line spectra from the IRAM 30m telescope; spectra of the first 4 galaxies are shown in Fig.~\ref{fig:spectra}. The spectra have been binned to a channel width of 30 \kms
  and the scale is in main beam temperature (mK). The blue line shows the best-fit Gaussian or double-peak function (fits are only performed on detected lines). The vertical dashed lines indicate the velocity range integrated over to obtain integrated intensities. {\em Top to bottom:}
  \two, \twt, \tho, \tht, HCN(1-0) and \hcopo. The \two and \twt data are from
  Paper IV and  \citet{welch03}. }
  \label{fig:otherspec}
\end{figure*}

\begin{figure*}
\begin{center}
\begin{tabular}{cccc}
\includegraphics[height=10cm]{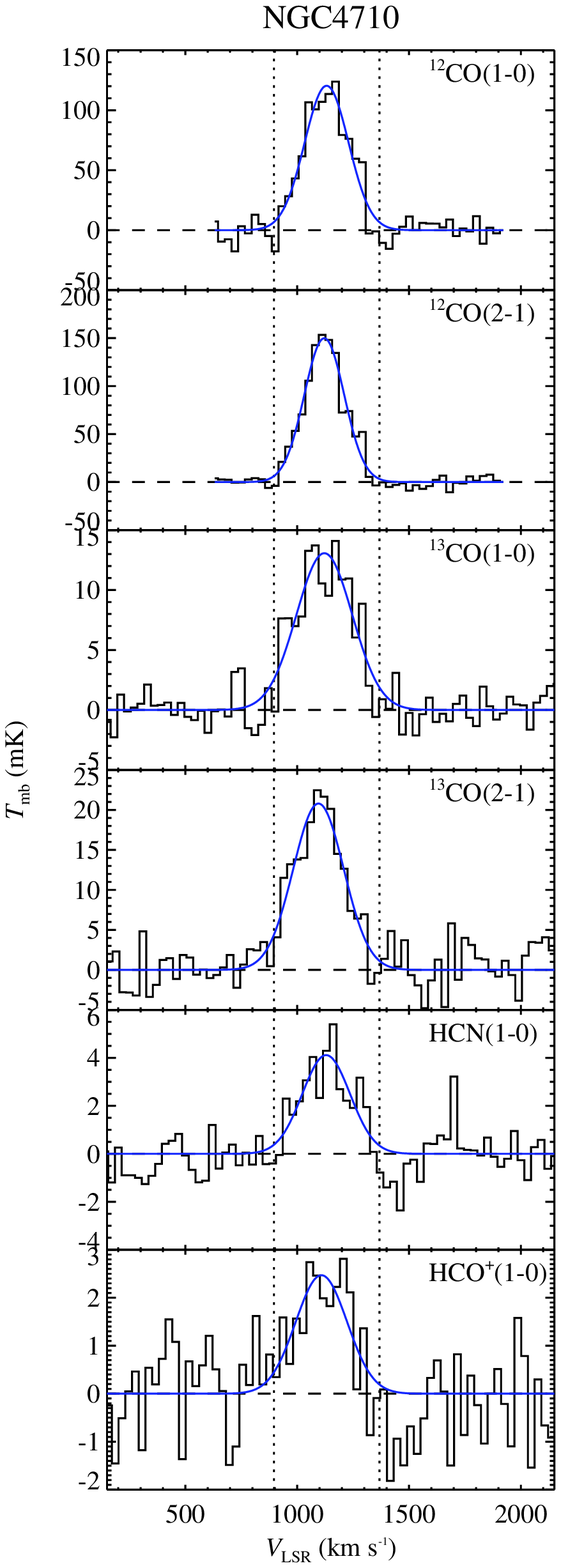} &
\includegraphics[height=10cm]{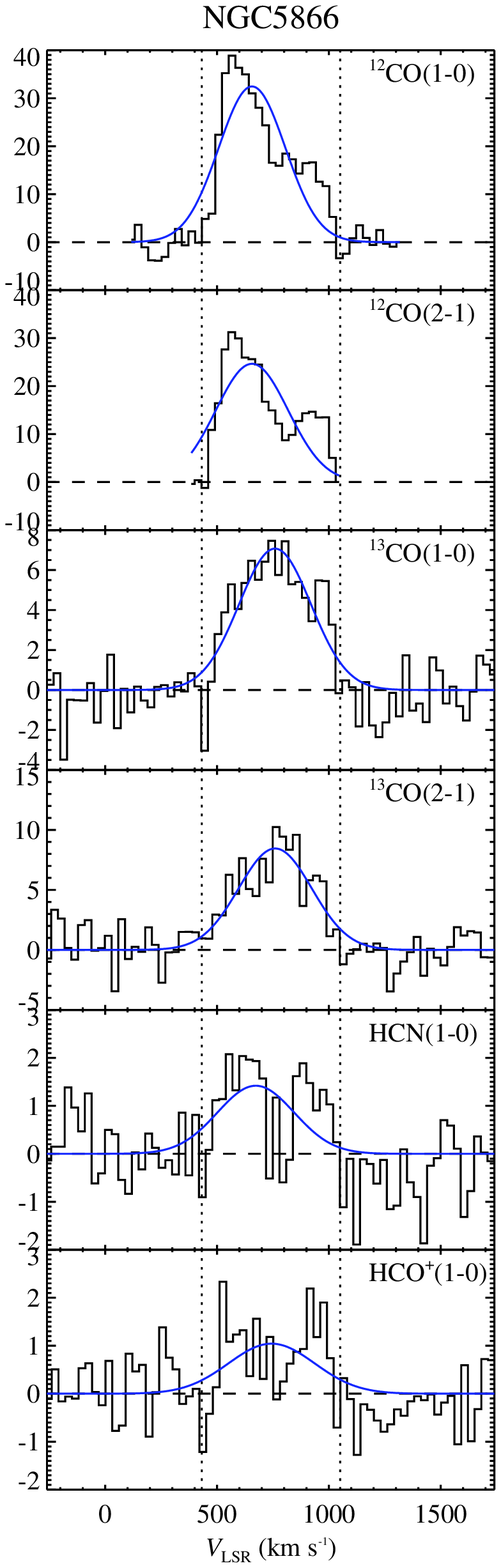} &
\includegraphics[height=10cm]{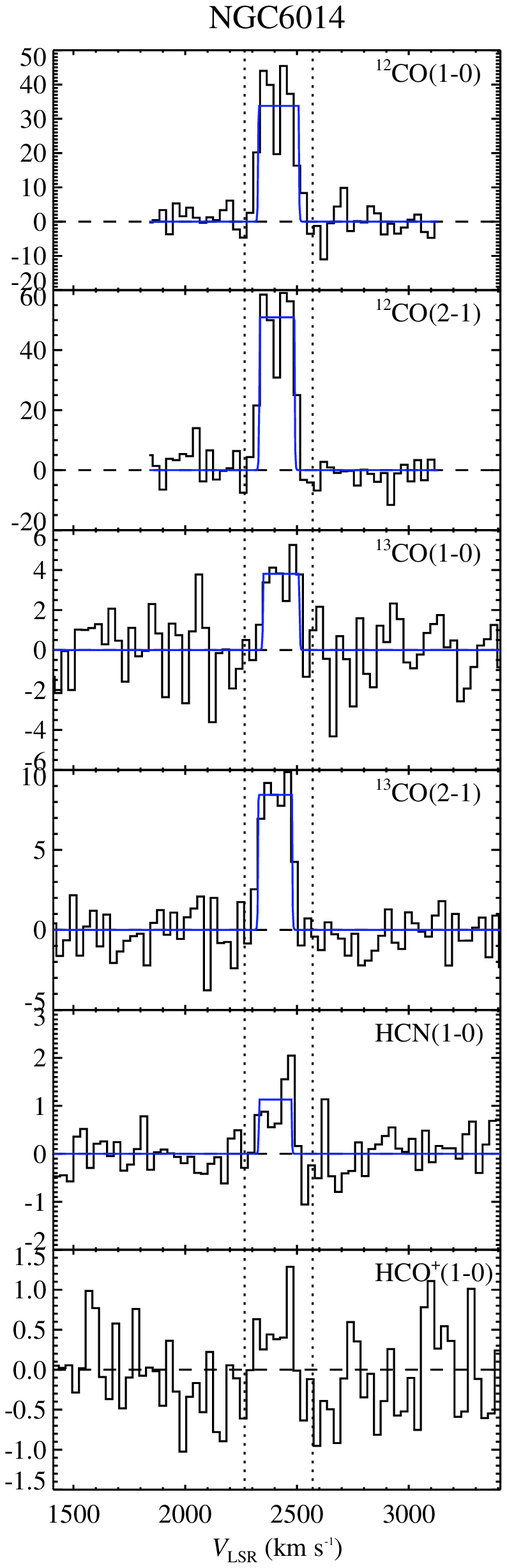} &
\includegraphics[height=10cm]{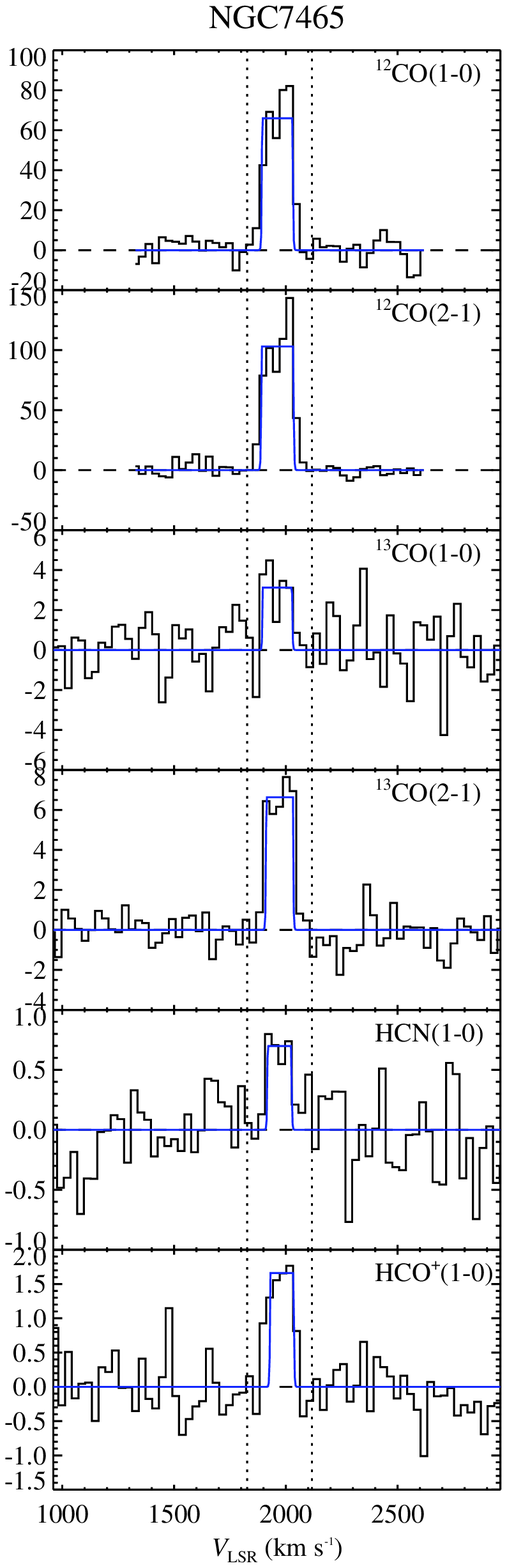}\\
\includegraphics[height=10cm]{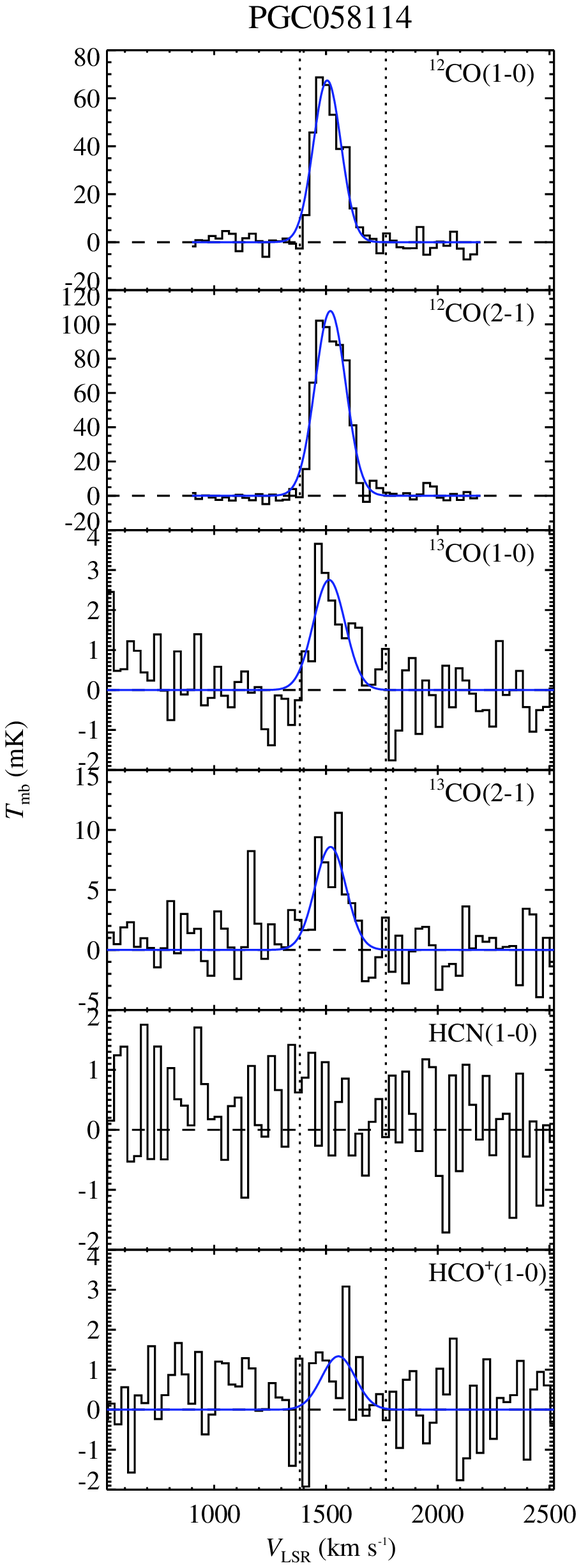} & 
\includegraphics[height=10cm]{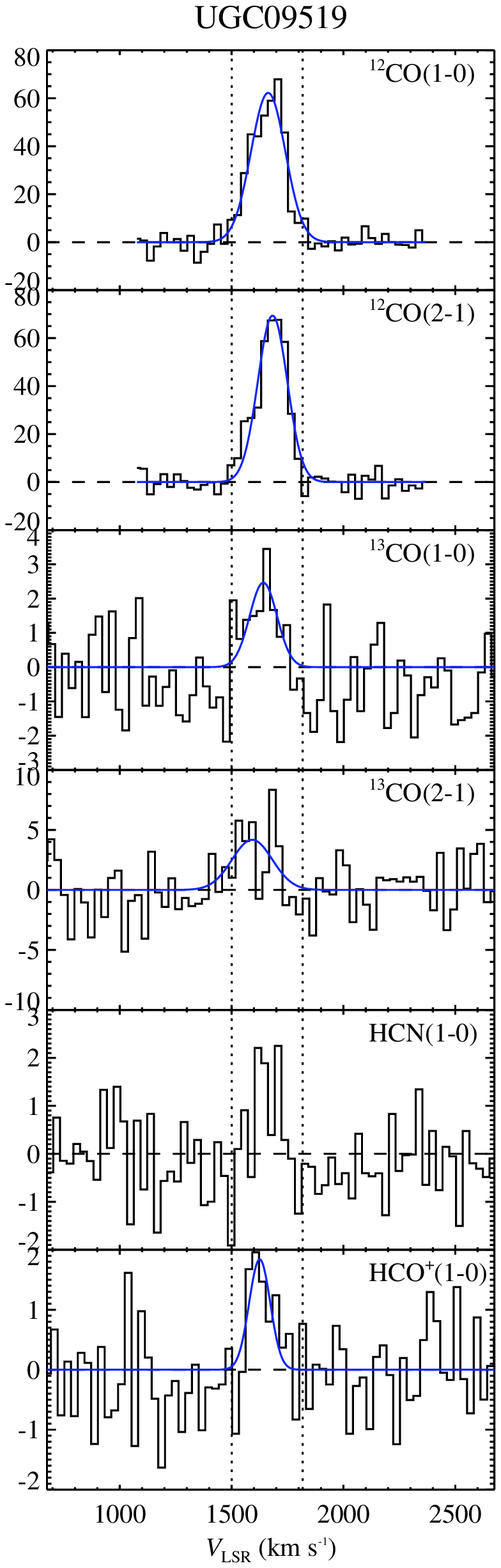} \\
\end{tabular}
\contcaption{}
\end{center}
\end{figure*}

\section{Test of line profile consistency}

Here we document how we test for line profile consistency. As the \two spectra generally have the highest signal-to-noise ratio, so we first fit a single Gaussian and a double-peaked profile to these lines. Our double-peak function is a parabola cut off symmetrically by a steeply declining exponential on either end:
\[f(v) = \left\{ 
\begin{array}{l l}
  \mathrm{T_{mb,w}}e^{-((\mathrm{v}_{0}-\mathrm{w}-v)/15 \mathrm{\,km\,s^{-1})}} & \quad v<\mathrm{v}_{0}-\mathrm{w}\\
  \mathrm{T_{mb,0}}+\mathrm{a}(v-\mathrm{v_0})^2 & \quad \mathrm{v}_{0}-\mathrm{w} \le v \le \mathrm{v}_{0}+\mathrm{w} \\
  \mathrm{T_{mb,w}}e^{-((v-\mathrm{v}_{0}-\mathrm{w})/15\mathrm{\,km\,s^{-1})}} & \quad v>\mathrm{v}_{0}+\mathrm{w}\\ 
  \end{array}. \right. \]
The parabola is a function of the velocity, $v$, and is centered at v$_{0}$, with a cutoff at v$_{0}\pm$w on either side. T$_{\mathrm{mb,0}}$ gives $T_{\mathrm{mb}}$ at  the central velocity, v$_{0}$, and the parameter `a' describes the steepness of the parabola ($a \ge 0$ in order to guarantee a double-peak or flat-topped shape). T$_{\mathrm{mb,w}}$=T$_{\mathrm{mb,0}}+a w^2$ is the peak height. Instead of a sharp cutoff from the peaks to zero, which leads to fitting problems, a steep exponential decay with a decay constant of 15 \kms is specified on both sides. 

The fits were performed with the Interactive Data Language (IDL) package MPFIT \citep{markwardt09}. The reduced $\chi^{2}$ values for the Gaussian and double-peak fits to the \two spectra are listed for each galaxy in Table~\ref{tab:chi2}. 7/18 galaxies are best fit by a single Gaussian, 6/18 by a true double-peak profile and 5/18 by a flat-topped profile ($a=0$). For most galaxies, this fit-based classification agrees with the by-eye classification in \citet[][hereafter Paper V]{davis11a}.

For all the other transitions of each galaxy, we constrain the type of fit to that which provides the lower $\chi^{2}$ value in the \two line. Uncertainties in the fit parameters are then determined by Monte Carlo simulations, adding Gaussian noise to the observed spectra and refitting 1000 times. The distributions of the resulting parameters are not necessarily Gaussian, so we use the equivalent to a 2$\sigma$ range, containing 95\% of the probability. The measured line profiles are considered consistent with each other when the uncertainties on all their parameters ($v_{0}$ and $\sigma$ for the Gaussian, $v_{0}$, $w$ and $a$ for the double-peak profile) overlap. We use the \two profile as a base and compare the other lines to it.

Table~\ref{tab:chi2} lists whether the line profile of each of the other five lines is consistent with that of \two (1 indicates consistency, 0 inconsistency). We note that consistency here is not that the intrinsic line profiles agree to within a certain measure, but instead that the observed line profiles do not rule out such agreement (i.e. are often too low signal-to-noise). 

\begin{table*}
\caption{Line profile fits and consistency.}
\begin{center}
\begin{tabular}{lcccccccc}
\hline
Name & Gaussian $\chi^{2}$ & Double-peak $\chi^{2}$ & Fit & \twt & \tho & \tht & \hcno &\hcopo \\
\hline
    IC0676 &    1.7 &    1.3 & F &      1  &      1  &      1  &      1  &      1 \\
    IC1024 &    2.9 &    2.2 & F &     0  &      1  &      1  &     ...  &     ... \\
   NGC1222 &    3.7 &    1.6 & D & 0  &      1  &      1  &     ...  &     ... \\
   NGC1266 &    2.0 &   78.7 & G & 0  &      0  &      0  &      0  &      0 \\
   NGC2764 &    1.8 &    1.8 & F & 1  &      1  &      1  &      1  &      1 \\
   NGC3032 &    2.1 &    0.9 & D & 1  &      0  &      1  &      1  &     ... \\
   NGC3607 &   11.4 &    3.0 & D & 0  &      0  &      0  &      0  &      1 \\
   NGC3665 &    4.0 &    2.4 & D & 1  &      1  &      1  &     ...  &     ... \\
   NGC4150 &    1.1 &    1.5 & G & 1  &      1  &      0  &     ...  &     ... \\
   NGC4459 &    3.3 &    2.2 & D &   1  &      1  &      1  &      1  &     ... \\
   NGC4526 &    5.8 &    2.1 & D & 0  &      0  &      0  &      0  &      0 \\
   NGC4694 &    1.0 &    1.4 & G &  0  &      1  &      1  &     ...  &     ... \\
   NGC4710 &    1.9 &    4.7 & G & 1  &      0  &      0  &      0  &      1 \\
   NGC5866 &    6.8 &    9.6 & G & 1  &      0  &      0  &      1  &      1 \\
   NGC6014 &    2.5 &    2.2 & F & 1  &      1  &      1  &      0  &     ... \\
   NGC7465 &    2.0 &    1.7 & F & 1  &     1 &      1  &     ...  &     ... \\
  UGC09519 &    1.6 &    6.7 & G & 0  &      1  &      1  &     ...  &     ... \\
 PGC058114 &    2.4 &    5.7 & G &  0  &      1  &      1  &     ...  &     ... \\

 \hline

\end{tabular}\\
\vspace{0.1cm}
Notes: The three types of fits denoted in the `Fit' column are a single Gaussian (G), a double-peaked profile (D) or a flat-topped profile (F). The final five columns indicate whether the line specified in the column heading has a profile consistent with the \two profile (1) or not (0). 

\end{center}
\label{tab:chi2}
\end{table*}%

\section{Derivation of beam correction factors}

A single-dish telescope measures an average surface brightness (usually expressed as a temperature) that results from the convolution of its main beam pattern with the sky brightness. The pattern of the main beam is roughly a Gaussian, with a half-power beam width (HPBW) that depends inversely on frequency. Thus, for a given telescope, a lower frequency line is measured over a larger sky region than a higher frequency line. For non-uniform sources, different intensities will be measured over differently sized beams, an effect that is often called `beam dilution' for central point sources or centrally concentrated sources.

Our data consist of the integrated intensities (i.e. brightness temperatures integrated over specific frequency or velocity ranges) of lines at different frequencies, therefore measured over different beam areas. For \hcno and \hcopo, the frequencies are so close that the difference in area is small ($\approx 1$\%) and their ratio can be computed without any correction. However, the difference is significant for all other line ratios. In the worst case, we compare the integrated intensity of \hcno and \hcopo in a 27.7 arcsec beam to the integrated intensity of \two in a 21.3 arcsec beam (a factor 1.7 in area). 

To predict and thus correct the effect of different beam sizes, the structure of the source must be known in the tracer used. Interferometric \two maps obtained for all sources as part of the \atlas project (Alatalo et al., in prep.) reveal the distribution of this line. Unfortunately, the other lines do not yet have interferometric data. Still, we can perform a first-order correction assuming that the distributions of the other lines are identical to that of \two, within a scaling factor. 

For each galaxy and for each transition, we first perform the weighted integration of the interferometric \two map using a Gaussian with appropriate HPBW as the weighting function.  The HPBWs we use are determined by the formula: HPBW$ = 2460\arcsec/[\nu/\mathrm{GHz}]$. This results in HPBWs of 27.7, 27.7, 22.3, 21.3, 11.2 and 10.7 arcsec for the \hcno, \hcopo, \tho, \two, \tht and \twt lines, respectively. Then we divide this spatially integrated flux by the beam area, obtaining an estimate of the integrated intensity of \two in each differently-sized beam. This derived intensity is written as $I_{\mathrm{HPBW}}^{\mathrm{int}}(^{12}$CO(1-0)), where the `int' denotes the interferometric origin and the HPBW is specified in arcseconds. With the assumption that other lines have spatial distributions identical to that of \two to within a scaling factor, we can then estimate their integrated intensities over different beam sizes. For example, to compute the estimated \hcno integrated intensity measured over the smaller \two beam:
\begin{equation}
I^{\prime}_{21.3\arcsec}(\mathrm {HCN}) = \frac{I_{21.3\arcsec}^{\mathrm{int}}(^{12}\mathrm{CO}(1\text{-}0))}{I_{27.7\arcsec}^{\mathrm{int}}(^{12}\mathrm{CO}(1\text{-}0))}\:I_{27.7\arcsec}(\mathrm {HCN}),
\end{equation}
where the prime indicates that this is an estimated, and not directly measured, integrated intensity. 

We list the correction factors required for the line ratios we study in Table~\ref{tab:corrfactors}. These have been applied to the ratios listed in Table~\ref{tab:ratios}. 

\begin{table*}
\caption{Integrated intensity correction factors.}
\begin{center}
\begin{tabular}{lccccc}
\hline
Name &   
 $\frac{I_{21.3\arcsec}^{\mathrm{int}}(^{12}\mathrm{CO}(1\text{-}0))}{I_{22.3\arcsec}^{\mathrm{int}}(^{12}\mathrm{CO}(1\text{-}0))}$ &
 $\frac{I_{10.7\arcsec}^{\mathrm{int}}(^{12}\mathrm{CO}(1\text{-}0))}{I_{11.2\arcsec}^{\mathrm{int}}(^{12}\mathrm{CO}(1\text{-}0))}$ &
$\frac{I_{21.3\arcsec}^{\mathrm{int}}(^{12}\mathrm{CO}(1\text{-}0))}{I_{27.7\arcsec}^{\mathrm{int}}(^{12}\mathrm{CO}(1\text{-}0))}$ & 
 $\frac{I_{22.3\arcsec}^{\mathrm{int}}(^{12}\mathrm{CO}(1\text{-}0))}{I_{27.7\arcsec}^{\mathrm{int}}(^{12}\mathrm{CO}(1\text{-}0))}$ &
 $\frac{I_{10.7\arcsec}^{\mathrm{int}}(^{12}\mathrm{CO}(1\text{-}0))}{I_{21.3\arcsec}^{\mathrm{int}}(^{12}\mathrm{CO}(1\text{-}0))}$  \\
 
 \\[-0.25cm]
\hline
 
 IC0676       &1.06&1.03&1.35&1.43&1.96\\
IC1024        &1.05&1.02&1.27&1.33&1.62\\
NGC1222    &1.05&1.02&1.27&1.33&1.63\\
NGC1266    &1.08&1.05&1.45&1.57&2.60\\
NGC2764    &1.06&1.05&1.34&1.43&2.28\\
NGC3032    &1.05&1.02&1.29&1.35&1.62\\
NGC3607    &1.06&1.03&1.32&1.39&1.92\\
NGC3665    &1.07&1.05&1.38&1.47&2.32\\
NGC4150    &1.05&1.04&1.30&1.38&1.97\\
NGC4459    &1.05&1.02&1.30&1.37&1.60\\
NGC4526    &1.05&1.03&1.28&1.34&1.71\\
NGC4694    &1.06&1.04&1.32&1.40&2.01\\
NGC4710    &1.05&1.02&1.31&1.38&1.68\\
NGC5866    &1.07&1.05&1.40&1.50&2.47\\
NGC6014    &1.07&1.04&1.38&1.47&2.29\\
NGC7465    &1.06&1.03&1.31&1.39&1.89\\
PGC058114 &1.07&1.07&1.40&1.50&2.24\\
UGC09519  &1.08&1.08&1.43&1.54&2.54\\

 \hline

\end{tabular}
\end{center}
\label{tab:corrfactors}
\end{table*}%

\label{lastpage}

\end{document}